\documentclass[twocolumn,amsmath,amssymb,superscriptaddress,aps,prb,floatfix,showpacs]{revtex4-1}

\usepackage{graphicx}
\usepackage{dcolumn}
\usepackage{bm}

\newcommand{\braket}[3]{\left\langle #1\;\vrule\; #2 \;\vrule\; #3\right\rangle}

\begin{document}
\title{Evolution of non-thermal phonon and electron populations in photo-excited germanium on picosecond timescales}
\author{F. Murphy-Armando}
\email{philip.murphy@tyndall.ie}
\affiliation{Tyndall National Institute, University College, Cork, Ireland}
\author{\'E. D. Murray}
\affiliation{Department of Physics and Department of Materials, Imperial College London, United Kingdom}
\author{I. Savi\'c}
\affiliation{Tyndall National Institute, University College, Cork, Ireland}
\author{M. Trigo}
\affiliation{PULSE Institute, SLAC National Accelerator Laboratory, Menlo Park, California 94025, USA}
\author{D. Reis}
\affiliation{PULSE Institute, SLAC National Accelerator Laboratory, Menlo Park, California 94025, USA}
\affiliation{Department of Photon Science and Applied Physics, Stanford University, Stanford, California 94305, USA}
\author{S. Fahy}
\affiliation{Tyndall National Institute, University College, Cork, Ireland}
\affiliation{Department of Physics, University College, Cork, Ireland}
\date{\today{}}

\begin{abstract}
We investigate from first-principles theory and experiment the generation of phonons on picosecond timescales and the relaxation of carriers in multiple conduction band valleys of photo-excited Ge by intervalley electron-phonon scattering.  
We provide a full description of the phonon and electron relaxation dynamics without adjustable parameters. 
Simulations of the time-evolution of phonon populations, based on first-principles band structure and electron-phonon and phonon-phonon matrix elements, are compared with data from time-resolved x-ray diffuse scattering experiments, performed at the LCLS x-ray free-electron laser facility, which measures the diffuse scattering intensity following photo-excitation by a 50 fs near-infrared optical pulse. 
Comparing calculations and measurements show that the intensity of the non-thermal x-ray diffuse scattering signal, that is observed to grow substantially near the L-point of the Brillouin zone over 3-5 ps, is due to phonons generated by scattering of carriers between the $\Delta$ and L valleys. 
Non-thermal phonon populations throughout the Brillouin zone are observed and simulated from first principles without adjustable parameters for times up to 10 ps. 
With inclusion of phonon decay through 3-phonon processes, the simulations also account for other non-thermal features observed in the x-ray diffuse scattering intensity, which are due to anharmonic phonon-phonon scattering of the phonons initially generated by electron-phonon scattering.
\end{abstract}

\maketitle

\section{Introduction}
The out-of-equilibrium behavior of electrons and atomic vibrations in materials underpin many natural phenomena, including phase changes, photoluminescence, photosynthesis, catalysis, and electronic and heat transport. 
However, most of our knowledge of the properties of materials is confined to the equilibrium regime.
As the advance of technology requires control of ever shorter time and space scales, a deeper understanding of the out-of-equilibrium dynamics of electrons and phonons becomes necessary.

Microscopic observation of out-of-equilibrium phenomena has become possible only recently, with the development of time resolved ARPES, X-ray FELs and time resolved electron diffraction.\cite{bovensiepen,reis,Popmintchev1287,PhysRevB.97.165416} These facilities allow us to measure the time evolution to equilibrium of momentum and energy resolved phonon and electron populations to better than 100 fs resolution.
These techniques provide an unprecedented way to benchmark the phenomena governing energy relaxation processes. In parallel to the experimental techniques we have an array of first principles electronic structure methods that yield the energy and momentum dependent electron-phonon coupling parameters.\cite{marini, murray,murray2,maldonado,ono,PhysRevB.98.144306,PhysRevLett.119.036803,PhysRevLett.119.136602,PhysRevB.98.134309,bernardi} The contribution of both the experimental and theoretical techniques allow us to interpret the measurements in terms of the microscopic phenomena.

In earlier work, Trigo {\it et al}\cite{trigo,PhysRevB.92.054303} used the diffuse scattering of femtosecond x-ray pulses to directly observe the time-dependence of the phonon distribution in photo-excited Ge arising from 
oscillations of the diffuse x-ray scattered intensity as the mean-square phonon displacement oscillates at twice the phonon frequency. Strong oscillatory signals were observed near the zone center\cite{trigo} and throughout the Brillouin zone\cite{PhysRevB.92.054303}, corresponding to squeezed acoustic phonon modes.

In this work, we investigate the diffuse x-ray scattering intensity near the Brillouin zone boundary, observed in Ge in the same scattering geometry as that in Ref. \onlinecite{PhysRevB.92.054303}. We observe a non-thermal gradual increase in the scattered intensity near the $L$ point, growing on a timescale of a few ps following the absorption of the optical pump pulse and saturating after that time. We use first-principles calculations and coupled rate equations to obtain the evolution of the non-equilibrium electron and phonon distribution after photo-excitation, in a similar way to Maldonado {\it et al.}\cite{maldonado} and Ono\cite{ono} for metals, and O'Mahony {\it et al.} for highly excited semimetals\cite{PhysRevLett.123.087401}. However, our approach is novel in that we require the detailed integration over small pockets of the Brillouin zone in the conduction band valleys. We accomplish this with the use of adaptive k-point grids rather than the usual uniform grid throughout the Brillouin zone. We use the calculated phonon population distribution to compute the x-ray diffuse scattering structure factor as a function of time to directly compare to our experiment.

The reason for the non-equilibrium phonon distribution near the $L$ point is quite subtle. The process of energy relaxation in this experiment can be divided into the following processes: i) The initial laser pulse inverts the electron population, chiefly at the $\Gamma$ conduction and valence band valleys in the Brillouin zone (50fs); ii) the non-equilibrium electron distribution relaxes its momentum and energy via inter-valley electron-phonon coupling (EPC) from the $\Gamma$ conduction band valley into the $L$ and $\Delta$ conduction band valleys (1ps); iii) further inter- and intra-valley EPC scattering occurs between the $L$ and $\Delta$ valleys (10ps); iv) the EPC generated non-equilibrium phonon distribution relaxes via 3-phonon processes across the Brillouin zone.

We find that the observed non-equilibrium phonon distribution near $L$ is due to process iii), which generates phonons with wave-vectors very close to $L$, namely the difference of wavevectors between the $L$ and $\Delta$ valleys. This process is notably slower than typical EPC timescales, taking several picoseconds. This is because: it starts later as it requires electron population to first accumulate in the $L$ and $\Delta$ valleys, it has a relatively weak EPC matrix element, and the $L-\Delta$ scattering rate also becomes weaker as the $\Delta$ valleys depopulate. 
Process ii) also generates phonons at the $L$ point. However, from our calculations we know that the polarization of these phonons is perpendicular to the x-ray scattering wavevector, and hence not visible due to the x-ray scattering selection rules (see eqs. \ref{eqs} and \ref{eqs:f} below). As known from the literature\cite{li,jap,tyuterev}, process ii) occurs within 180 fs, rather than over several ps. Process iv) is observed as a slow diffuse increase of intensity in other areas of the Brillouin zone.




The paper is organized as follows. We start with a discussion of the experimental and theoretical observations in section \ref{sres}. We give the details of the experimental setup in Section \ref{sexp}. Details of the electron-phonon and the phonon-phonon scattering calculations are given in sections \ref{selph} and \ref{sphph}, respectively. The Appendix contains details of the integration of the electron-phonon and phonon-phonon coupling.

\section{Results}\label{sres}

The change of the diffuse x-ray scattering intensity after photo-excitation $I({\bf \Delta k})$ is given by
\begin{equation}
\Delta I({\bf \Delta k},t) = I({\bf \Delta k},t)-I({\bf \Delta k},t<0),
\end{equation}
where the intensity at time $t$ and wave-vector ${\bf \Delta k}$ is given by\cite{xray}
\begin{equation}\label{eqs}
I({\bf \Delta k},t)=\sum_\eta \frac{\hbar}{2 \omega_{{\bf q},\eta}} \left(N_{{\bf q},\eta}(t)+\frac{1}{2}\right) \left|F_\eta ({\bf \Delta k})   \right|^2  
\end{equation}
with
\begin{equation}\label{eqs:f}
F_\eta({\bf \Delta k})=\sum_{\alpha} \frac{1}{\mu_{\alpha}} e^{-M_{\alpha}} e^{-i {\bf G_{\Delta k}.r_\alpha}}{\bf \Delta k}\cdot {\bf u}_{\alpha ,{\bf q}}^\eta
\end{equation}
where $N_{{\bf q},\eta}$ and ${\bf u}^\eta_{\bf\alpha,q}$ are the phonon occupations and eigenmodes for wave-vector ${\bf q}$, polarization $\eta$ and atom $\alpha$ in the primitive cell, respectively. $\mu_\alpha$ and $M_\alpha$ are the mass and the Debye-Waller factor of the atom at $\alpha$, respectively. The factor ${\bf G_{\Delta k}}$ corresponds to the the reciprocal lattice vector at each ${\bf \Delta k=G_{\Delta k}+q}$ for a phonon wave-vector ${\bf q}$ in the first Brillouin zone (000).

\begin{figure}
\includegraphics[width=3.4in] {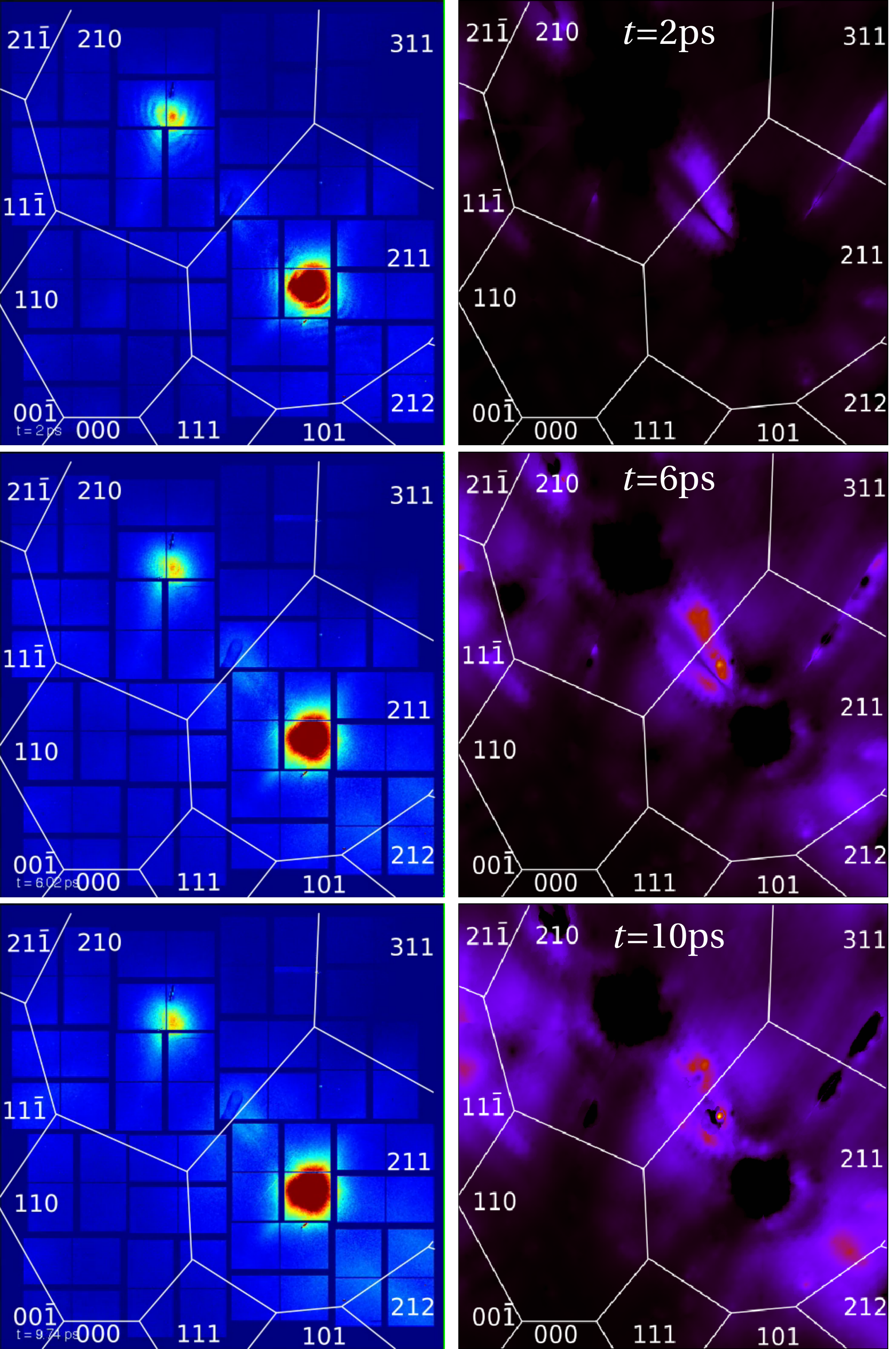}
\caption{\label{fig1exp} Experimental (left) and calculated (right) x-ray intensity at 2, 6 and 10 ps after the pump pulse. Points corresponding with the Bragg condition are seen as very bright spots (left) and are ignored in the calculation. Note the fringes near the zone centers due to the squeezed phonons on the top-left panel. }
\end{figure}

Fig. \ref{fig1exp} shows the measured and calculated first 10 ps of the diffuse x-ray intensity difference across a section of the Brillouin zone after excitation with a 50 ps, 1.5 eV laser pulse. The two bright spots at the center of zones (210) and (211) in the figures on the left correspond to the acoustic modes near the zone center in the experiment, which are excited directly by the pump through a sudden change in the forces leading to phonon squeezing \cite{trigo}. This process is not the focus of this work and is not accounted for by the calculations. 

\begin{figure}
\includegraphics[width=3.4in] {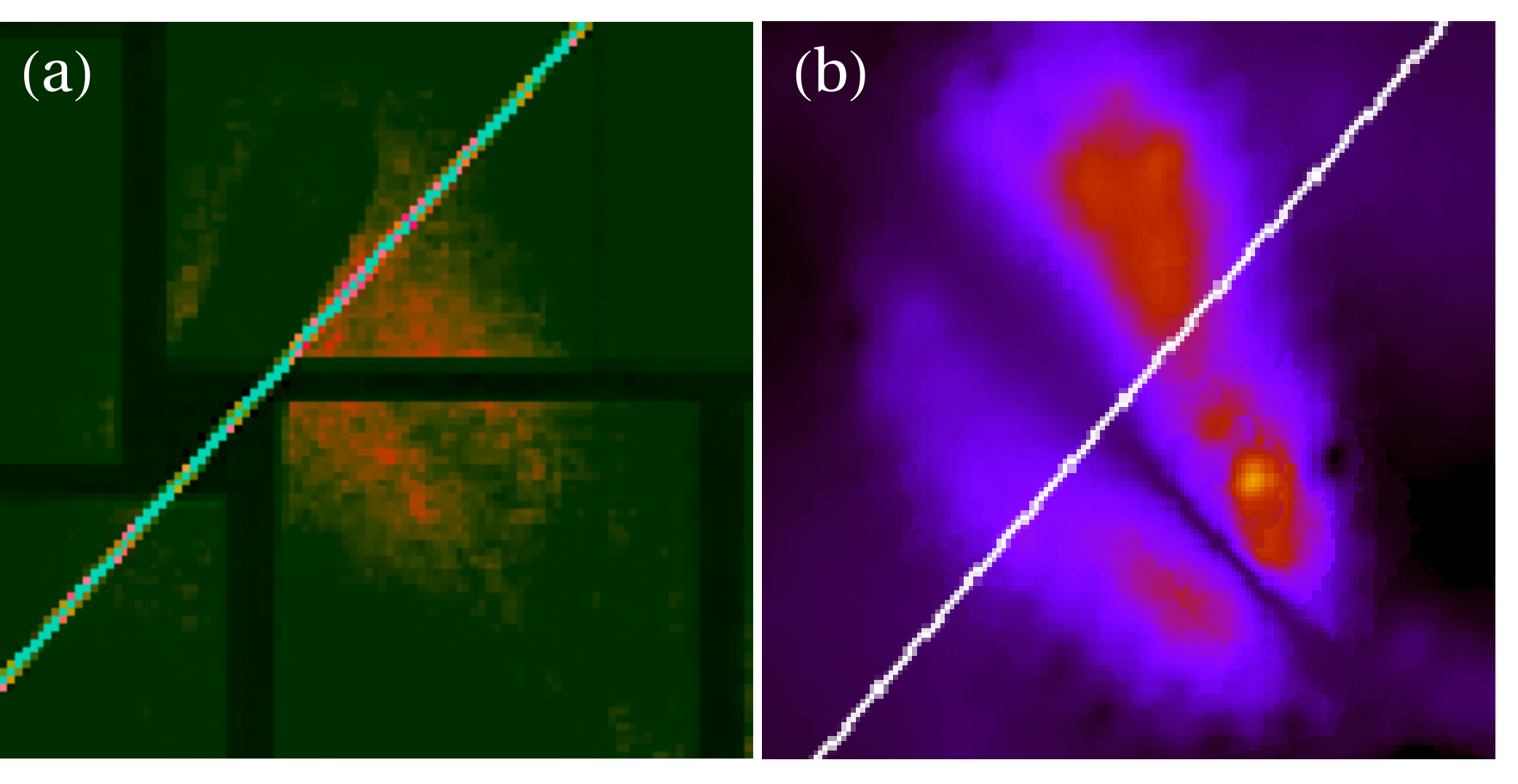}
\caption{\label{zexpvth} Detail of the experimental (left) and calculated (right) x-ray intensity at the zone edge near the $L$ point at 6 ps after the pump pulse (colors changed for highlight). The ``coffee bean" shape of the x-ray intensity, captured both in theory and experiment, is due to phonons generated by electrons scattering between the $L$ and $\Delta$ electronic valleys. }
\end{figure}

The interesting feature with a ``coffee bean" shape of the signal seen both in the experiment and calculations (see detail of Fig. \ref{fig1exp} in Fig. \ref{zexpvth}) and the slow increase in intensity at the $L$ point, in the boundary of zones (210) and (211) are due to electron-phonon inter-valley scattering. Our calculations determine this increase corresponds to the non-equilibrium increase in population of the $L\Delta,TA$ phonon generated by EPC, rather than phonons generated at $L$. The $L\Delta,TA$ phonon wave-vector is adjacent to the $L$ point, as shown in Fig. \ref{GeBZ}, and therefore the signal from this phonon appears in the BZ boundary region near the $L$ point. In fact, the product ${\bf \Delta k}\cdot {\bf u}_{\alpha ,{\bf q}}^\eta$ in Eq. \ref{eqs:f} forbids the appearance of the $L,LO$ phonon generated by $\Gamma$ to $L$ inter-valley scattering in this slice of the Brillouin zone. Furthermore, unlike the $L,LO$ phonon, we find a close correspondence between the measured and calculated time evolution of the heating of the $L\Delta,TA$ phonon.
Of particular note is the shape of this bright area at the $L$ point. In both theory and experiment it resembles the shape of a coffee bean, with the line joining the zone centers showing no intensity. This effect is due to a combination of two factors. The overall extent of the area is given by the extent of the quasi-Fermi surface at the $L$ and $\Delta$ valleys. The line of low intensity across the middle of the bean is due to the diffuse x-ray selection rule given by ${\bf \Delta k}\cdot {\bf u}_{\alpha ,{\bf q}}^\eta$. The latter is strictly nil along the line joining the zone centers only for the polarizations of the $\Delta L$,$TA$ phonons generated by EPC for this experimental geometry. This is the only polarization visible in that region of the BZ. Therefore, the non-equilibrium phonon occupations reveal details of the electronic distribution, band structure and electron-phonon coupling.

Concurrent with the process discussed, anharmonicity works to thermalise this highly non-equilibrium distribution and results in a slowly increasing background over the entire BZ, a signature that the system is evolving towards equilibrum. 

Finally, Fig. \ref{fig1exp} (left) shows fringes near the zone centers, especially in the first 3 ps. These are due to squeezed coherent acoustic phonons,\cite{trigo} and are not included in the model of this work.

\begin{figure}
\includegraphics[width=3.4in] {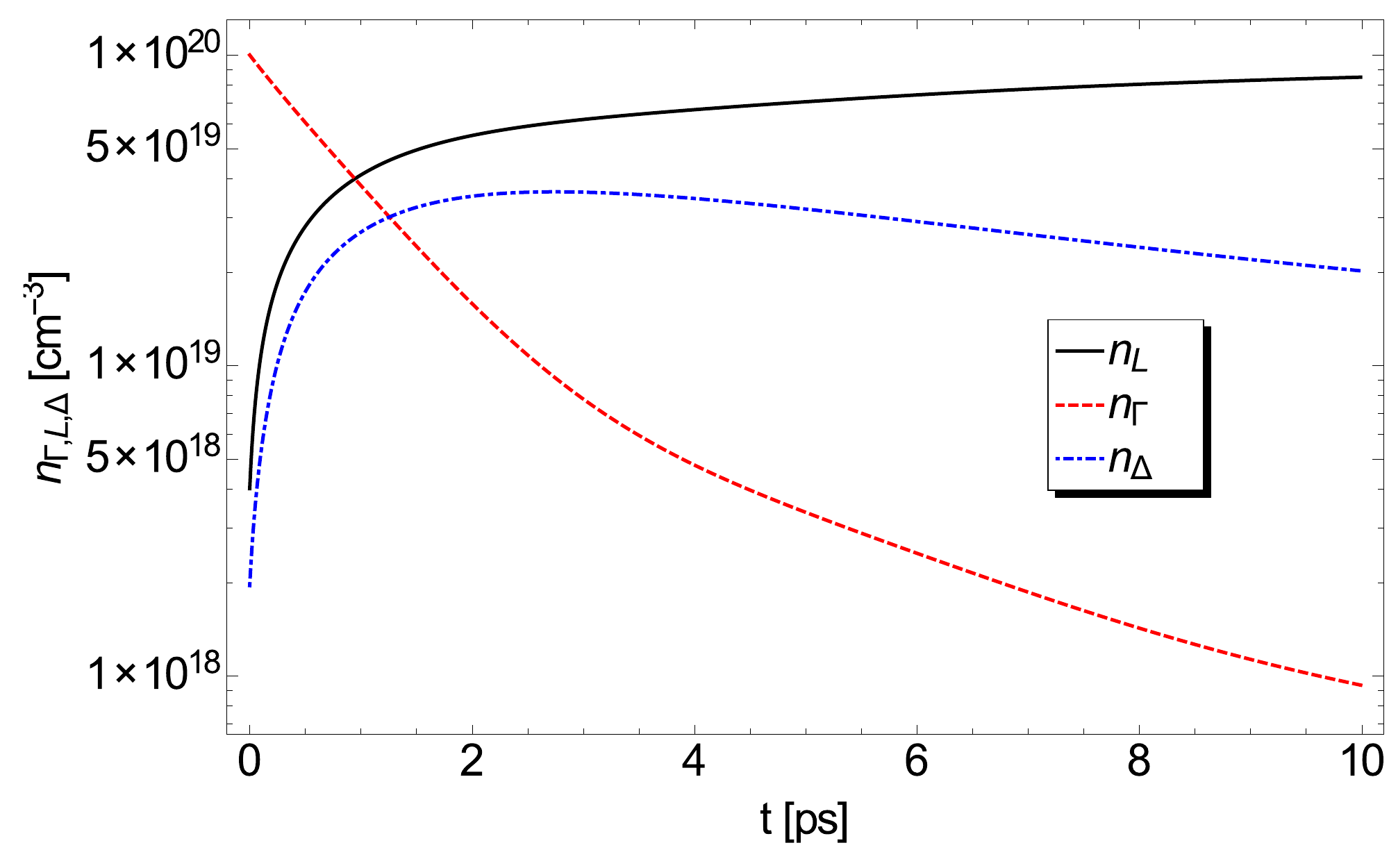}
\caption{\label{nelvst} Calculated evolution of the carrier concentration of the $\Gamma$, $L$ and $\Delta$ conduction band valleys after photo-excitation.}
\end{figure}

The calculated time evolution after photo-excitation of the carrier density at the $\Gamma$, $L$ and $\Delta$ conduction bands is shown in Fig. \ref{nelvst}. The initial photo-excited 
population at the $\Gamma$ valley quickly scatters into the  $L$ and $\Delta$ valleys within the first ps. Later, as the electrons lose energy to the lattice the carriers at the $\Delta$ valley scatter 
into the $L$ valley. Meanwhile, inter-valley $L\rightarrow L'$ and $\Delta\rightarrow \Delta'$ scattering occurs. Most of the energy of the carriers is lost to the lattice via intra-valley $L\rightarrow L$ scattering through the $\Gamma$ $LO$ phonon.

\begin{figure}
\includegraphics[width=3.4in] {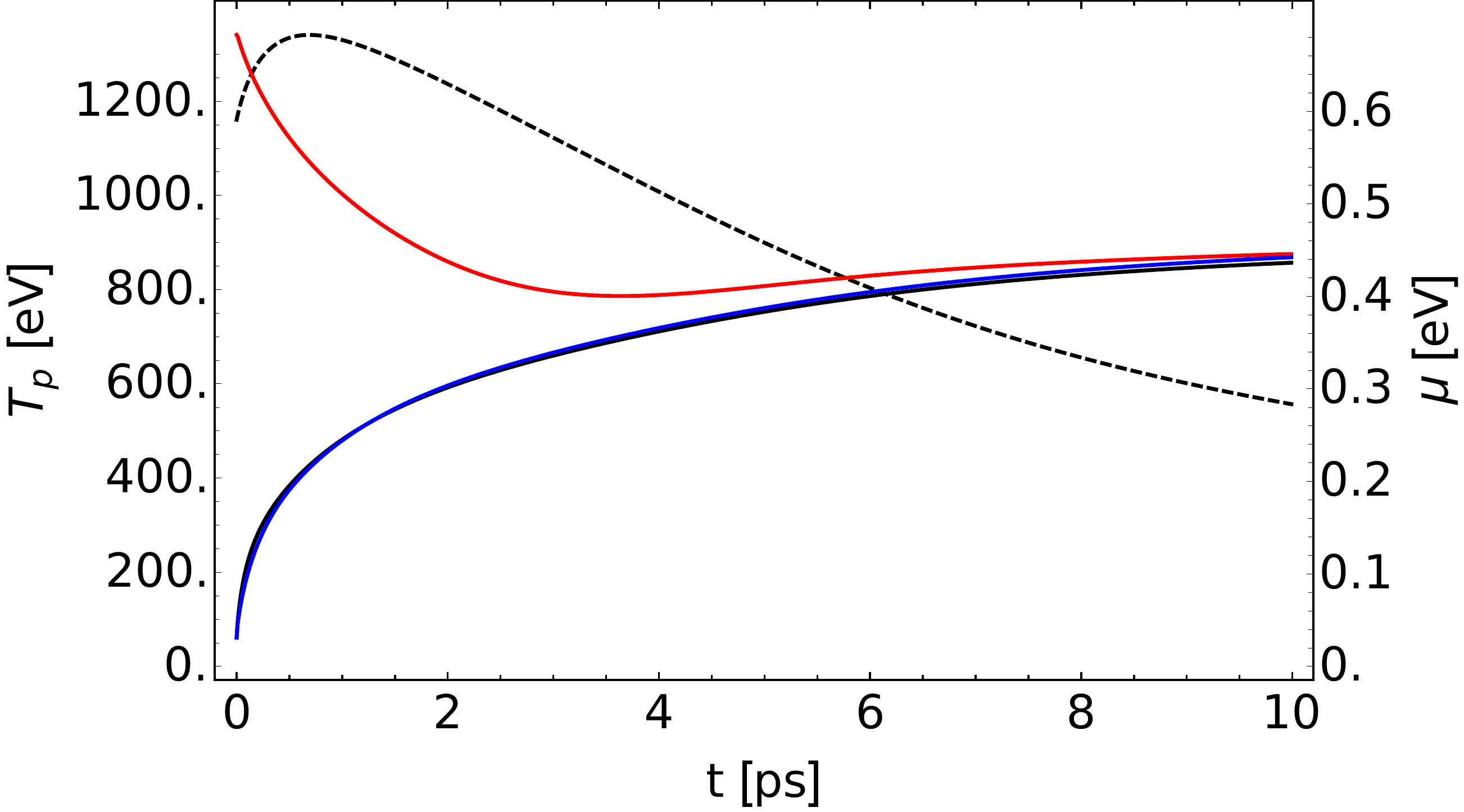}
\caption{\label{tpvst} Calculated evolution of the plasma temperature $T_p$ (dashed) and chemical potential of the $\Gamma$ (red), $L$ (black) and $\Delta$ (blue) conduction band valleys after photo-excitation.}
\end{figure}

The evolution of the plasma temperature and chemical potentials of the conduction bands are shown in Fig. \ref{tpvst}. The initial rise in temperature is due to the kinetic energy gained by the carriers transferring from  the $\Gamma$ valley to the lower $L$ valley. After photo-excitation, the carrier population of the $\Gamma$ conduction band takes 10ps to regain equilibrium with the $L$ and $\Delta$ bands. However, at 10 ps the plasma is still out of equilibrium with the lattice.

\begin{figure}
\includegraphics[width=3.4in] {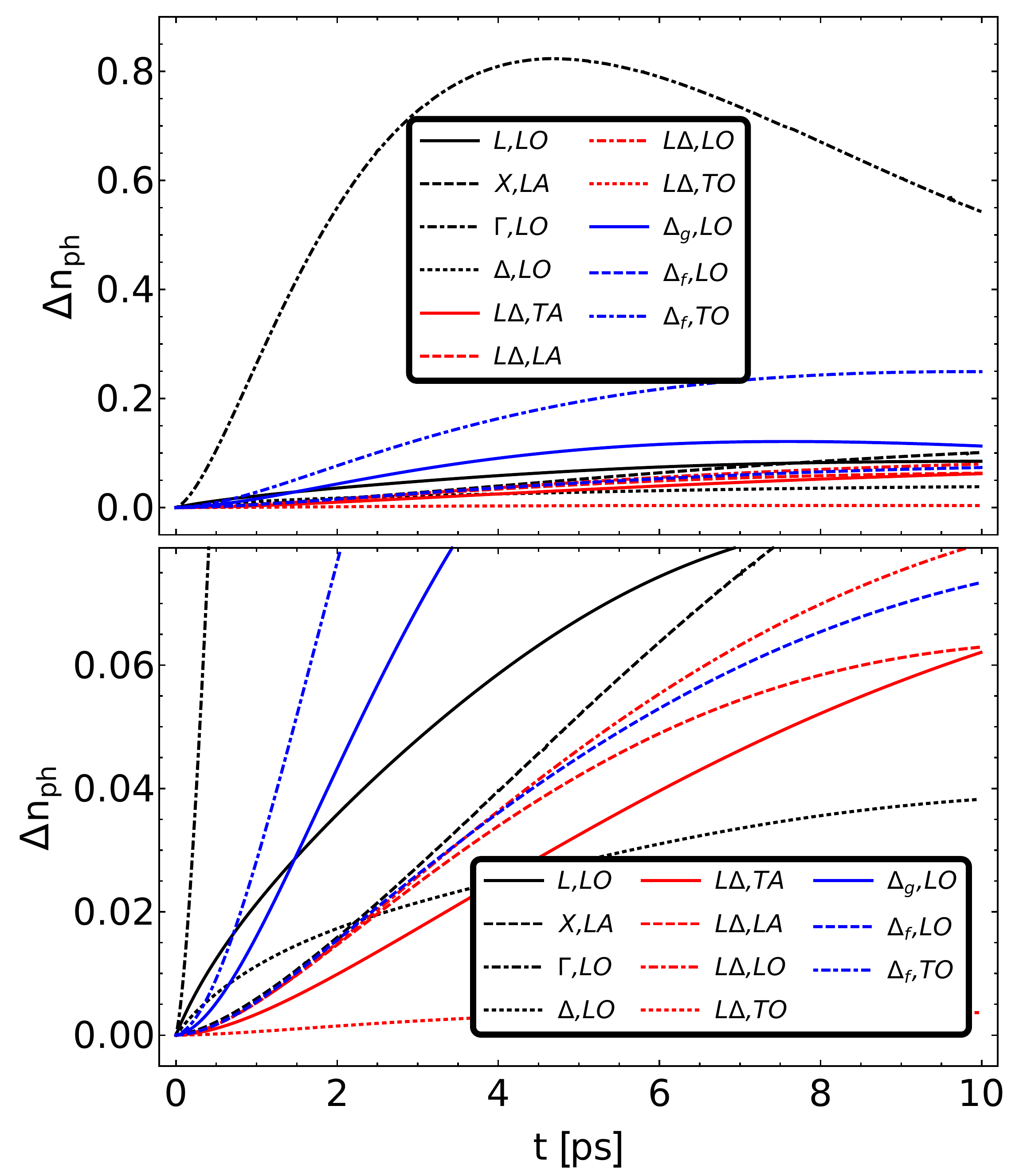}

\caption{\label{dnvst} Calculated change in the phonon population generated by EPC and anharmonic phonon-phonon decay after photo-excitation. Bottom: detail of top figure for the phonons with weaker rate of change. In red: $\Delta L$ phonons visible in the x-ray diffuse scattering image in Fig. \ref{fig1exp}.}
\end{figure}

The change in the phonon population generated by the inter-valley processes described above are shown in Fig. \ref{dnvst}. Evident are the initial increase in the $L$ and $\Delta$ phonons generated by the $\Gamma\rightarrow L$ and $\Gamma\rightarrow \Delta$ transitions in the first ps, characterized by a concave initial change in phonon population. Conversely, the $X$ phonon generated by the $L\rightarrow L'$ transition steadily increases its population, being one of two transitions that remain after carriers have cooled down into the $L$ valley. The other is the $L\rightarrow L$ intra-valley transition that generates the optical phonon at $\Gamma$, and is responsible of most of the heating of the lattice.

Other transitions that begin much later after the initial depopulation of the $\Gamma$ band generate the $\Delta-L$ $LA$, $TA$, $LO$ and $TO$ phonons. These increase very slowly over the first 10ps, due to a weak EPC and fast depopulation of the $\Delta$ valley. As we shall see below, these phonons are the only ones visible in our experiment.

The strong decay of the $\Gamma$ $LO$ phonon is due to the re-absorption of phonons by the cooling electrons, rather than by anharmonic decay. The LO phonon absorbs most of the electron heat, and reaches a temperature very close to the electron temperature. While not overshooting, this is a similar case of that observed in GaAs\cite{PhysRevB.43.4158}, where the temperature of the polar phonon overshoots the plasma temperature and has a similar cooling rate.





\section{Experimental setup}\label{sexp}

The measurements of the diffuse x-ray scattering in Ge were performed at the Linac Coherent Light Source (LCLS) x-ray free electron laser using 50 fs, 1.55 eV pump pulses and 50 fs, 10 keV x-ray probe pulses. The experimental data was acquired under the same conditions as in Ref. \onlinecite{PhysRevB.92.054303}.

\section{Theoretical modeling}
We use coupled Boltzmann rate equation to calculate the electron occupation at the $\Gamma$, $L$ and $\Delta$ conduction band valleys in Ge and the phonons generated by EPC after photo-excitation. The photo-excited electrons initially populate the $\Gamma$ conduction band valley, and later scatter into other valleys only via electron-phonon coupling. We assume that thermalisation via electron-electron interaction occurs at each valley within 100 fs. \cite{goldman} 
Therefore each valley contains a Fermi-Dirac distribution with a distinct chemical potential. Except for the initial depopulation of the $\Gamma$ valley, the relative changes in valley population are small during the typical thermalization time (estimated to be ~100 fs) for the carriers within each valley, so we expect the latter approximation to be valid.
 We have ignored the effects of the valence band on the generation of phonons, as they only generate phonons with small wave-vector around $\Gamma$.

\subsection{Electron-phonon scattering}\label{selph}
In this work we are chiefly concerned with the phonons produced by inter-valley EPC. The scattering rate of an electron at the first conduction band in a state of momentum $\bf k$ in valley $i$ to a state of momentum $\bf k+q$ in valley $j$ by a phonon of momentum $\bf q$ and branch $\eta$ is given by Fermi's Golden Rule,
\begin{equation}
R_{\bf k, k+q}^{\pm,\eta,i,j}=\frac{2\pi}{\hbar}\left|M_{\bf k, k+q}^{\pm,\eta,i,j}\right|^2 \delta(E_{\bf k+q}^j-E_{\bf k}^i\pm\hbar \omega_{\bf q}^\eta).
\end{equation}
where the matrix element can be expressed as a function of inter-valley deformation potentials $M_{\bf k, k+q}^{\pm,\eta,i,j}=D_{\bf k, k+q}^{\eta,i,j} \sqrt{\left( N_{\bf q}^\eta(t)\pm\frac{1}{2}+\frac{1}{2}\right)\frac{\hbar}{2 \rho V \omega_{\bf q}^\eta} f^i_{\bf k} \left(1- f^j_{\bf k+q}\right)}$. Here, $D_{\bf k, k+q}^{\eta,i,j}$ is the deformation potential for a phonon of crystal momentum $\bf q$, branch index $\eta$, occupation $N_{\bf q}$ and frequency $\omega_{\bf q}^\eta$ scattering an electron with momentum $\bf k$ in valley $i$ and occupation $f^i_{\bf k}$ to one with momentum $\bf k+q$ in valley $j$ and occupation $f^j_{\bf k+q}$. We omit the electron band index as we only consider the first conduction band. $\rho$ is the mass density and $V$ the volume of the system in which the wavefunctions are normalised (typically the primitive cell).

The rate of change of the carrier population $n_i(t)$ of valley $i$ and of phonon population $N_{\bf q}^\eta (t)$ of momentum $\bf q$ and branch $\eta$ due to electron-phonon and phonon-phonon scattering are given by
\begin{eqnarray}
&\dot{n}_i(t)&=\sum_j (R_{j,i}^++R_{j,i}^--R_{i,j}^+-R_{i,j}^-),\\
&\dot{N}_{\bf q}^\eta(t)&=\dot{N}_{\bf q}^\eta(t)|_{el-ph}+\dot{N}_{\bf q}^\eta(t)|_{ph-ph},\label{ndot}\\
&\dot{N}_{\bf q}^\eta(t)|_{el-ph}&=\sum_{ij} \left(\Gamma_{\bf q}^{+,\eta,i,j}-\Gamma_{\bf q}^{-,\eta,i,j}\right),
\end{eqnarray}
where $R_{i,j}^\pm$ is the electron-phonon scattering rate from valley $i$ to $j$ (see eq. \ref{rij} below), and $+,-$ accounts for emission and absorption of a phonon, respectively. 
The rate of change of the phonon distribution $\Gamma_{\bf q}^{\pm,\eta,i,j}$ due to the EPC is given by
%
\begin{eqnarray}\label{rq}
&\Gamma_{\bf q}^{\pm,\eta,i,j}&=\sum_{\bf k} R_{\bf k, k+q}^{\pm,\eta,i,j}=\nonumber \\
&=&\frac{2\pi}{\hbar}\left|D_{\bf q_0}^{\eta,i,j}\right|^2  \left( N_{\bf q}^\eta\pm\frac{1}{2}+\frac{1}{2}\right)\frac{\hbar}{2 \rho \omega^\eta_{\bf q_0}} \frac{1}{\left(2\pi\right)^3} \times \nonumber \\
&\times & \int_{BZ} {\bf dk} f^i_{\bf k}\left(1- f^j_{\bf k+q}\right) \delta(E_{\bf k+q}^j-E_{\bf k}^i\pm\hbar \omega^\eta_{\bf q_0})
\end{eqnarray}

In eq. \ref{rq} we have exploited the fact that the factor $f_{\bf k}\left(1- f_{\bf k+q}\right)$ limits the momentum $\bf q$ of phonons that can participate in the scattering to just a few in the vicinity of the momentum difference between two valleys. Within that phase space, we expect little change in $D_{\bf k,k+q}$, and $\omega_{\bf q}$ with $\bf q$ around select $\bf q_0$, where ${\bf q_0}={\bf k}_i-{\bf k}_j$ is a wave-vector  joining the energy minimum of two conduction band valleys $i$ and $j$. Therefore $D_{\bf q_0}^{\eta,i,j}$ is the electron-phonon coupling constant (inter-valley deformation potential, given in Table \ref{param}) for phonon $\bf q_0$ that joins valleys $i$ and $j$ with frequency $\omega_{\bf q_0}$.

In the present work, we will only be following the total population at each valley. Therefore, the scattering rate from valley $i$ to valley $j$ is given by
\begin{equation}\label{rij}
R_{i,j}^\pm= 
\frac{1}{V} \sum_{\bf q,\eta} \Gamma_{\bf q}^{\pm,\eta,i,j}
=\sum_{\eta}\frac{1}{(2 \pi)^3}\int_{BZ} \Gamma_{\bf q}^{\pm,\eta,i,j}{\bf dq} 
\end{equation}

We account for anharmonic phonon decay in the populations affecting the inter-valley EPC by the introduction of the second term in eq. \ref{ndot}. $\dot{N}_{\bf q}^\eta(t)|_{ph-ph}$ is the rate of change of phonons due to anharmonic phonon-phonon scattering. The details of how this rate updates the phonon populations are given in the following section. In the calculation of the electronic inter-valley scattering rates we only need to update the population of the phonons that are involved in electron inter-valley scattering,
\begin{equation}
\dot{N}_{\bf q_0}^\eta(t)|_{ph-ph}=-\frac{(N_{\bf q_0}^\eta(t) - N_{\bf q_0}^\eta(0))}{\tau_{\bf q_0}^\eta}
\end{equation}
where $\tau_{\bf q}^\eta$ is the anharmonic decay time, given in Table \ref{param} for the phonons at $\bf q_0$ involved in inter-valley scattering, and $N_{\bf q}^\eta(0)$ is the phonon occupation at $t=0$, equal to the equilibrium distribution at room temperature.


The phonon occupations in Eq. \ref{rq} are updated at each time step by those given by Eq. \ref{ndot} in a grid of $20\times20\times20$ $\bf q$ points. The integration in Eq. \ref{rq} is performed on a very small volume of the Brillouin zone that varies strongly with $\bf q$. This requires very dense $\bf k$- and $\bf q$-point grids. To make the integration more efficient, we only perform the integration where the conservation of energy condition is met, and where integrands are non-vanishing, using adaptive grids. Details of these integrals are given in Appendix \ref{ap:int}.

\begin{figure}
\includegraphics[width=3.2in] {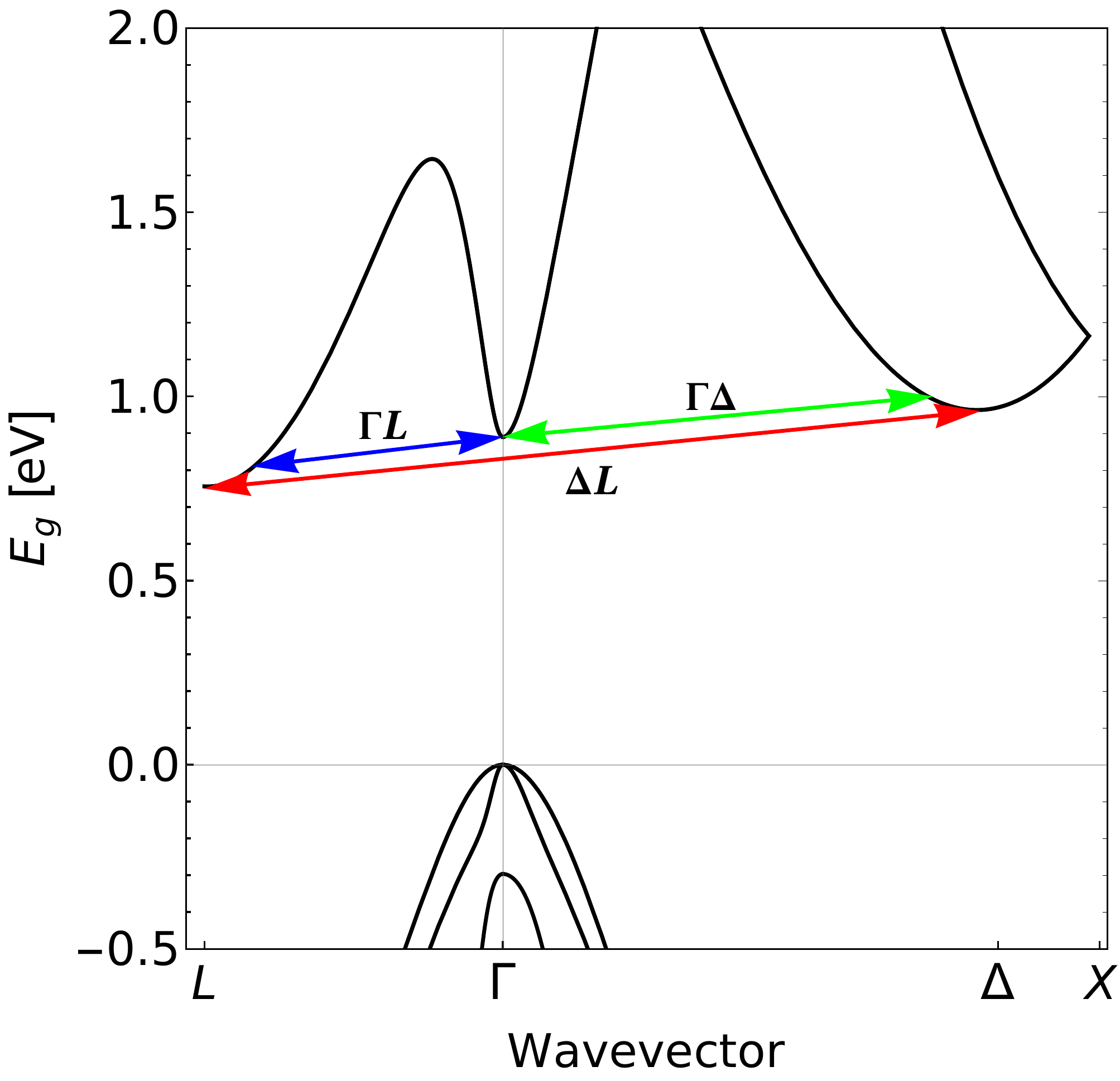}
\caption{\label{ebs} Electronic band structure of Ge. The arrows show the transitions between the $\Gamma$, $\Delta$ and $L$ valleys that generate the non-equilibrium phonon population.}
\end{figure}

The electronic dispersion used in the integration of the rate equations is obtained using the 30 band k.p approach of Rideau {\it et al}\cite{kdp} (shown in Fig. \ref{ebs}). The electron-phonon matrix elements and phonon lifetimes are calculated from first principles. The electron-phonon results have been previously calculated in Refs. \onlinecite{prb2,jap,jap2,prberratum}, and are recalculated here on a denser $\Gamma$ centered grid of $10^3$ k-points, and plane wave energy cut-off of 60Ha, with the aim of getting points closer to the $\Delta$ point. The electron phonon matrix elements are calculated using Abinit\cite{abinit} with Hartwigsen-Goedecker-Hutter\cite{hgh} pseudo-potentials. In our electron-phonon coupling calculations we only include phonons that are likely to be seen in the x-ray intensity experiment, and those that relax the energy, viz. optical phonons and phonons with large wave-vector. These phonons are generated by inter-valley and intra-valley optical phonon scattering via EPC in the conduction band. The details for the calculation of the phonon relaxation time by anharmonic decay are treated in section \ref{sphph}.

\begin{figure}
\includegraphics[width=3.2in] {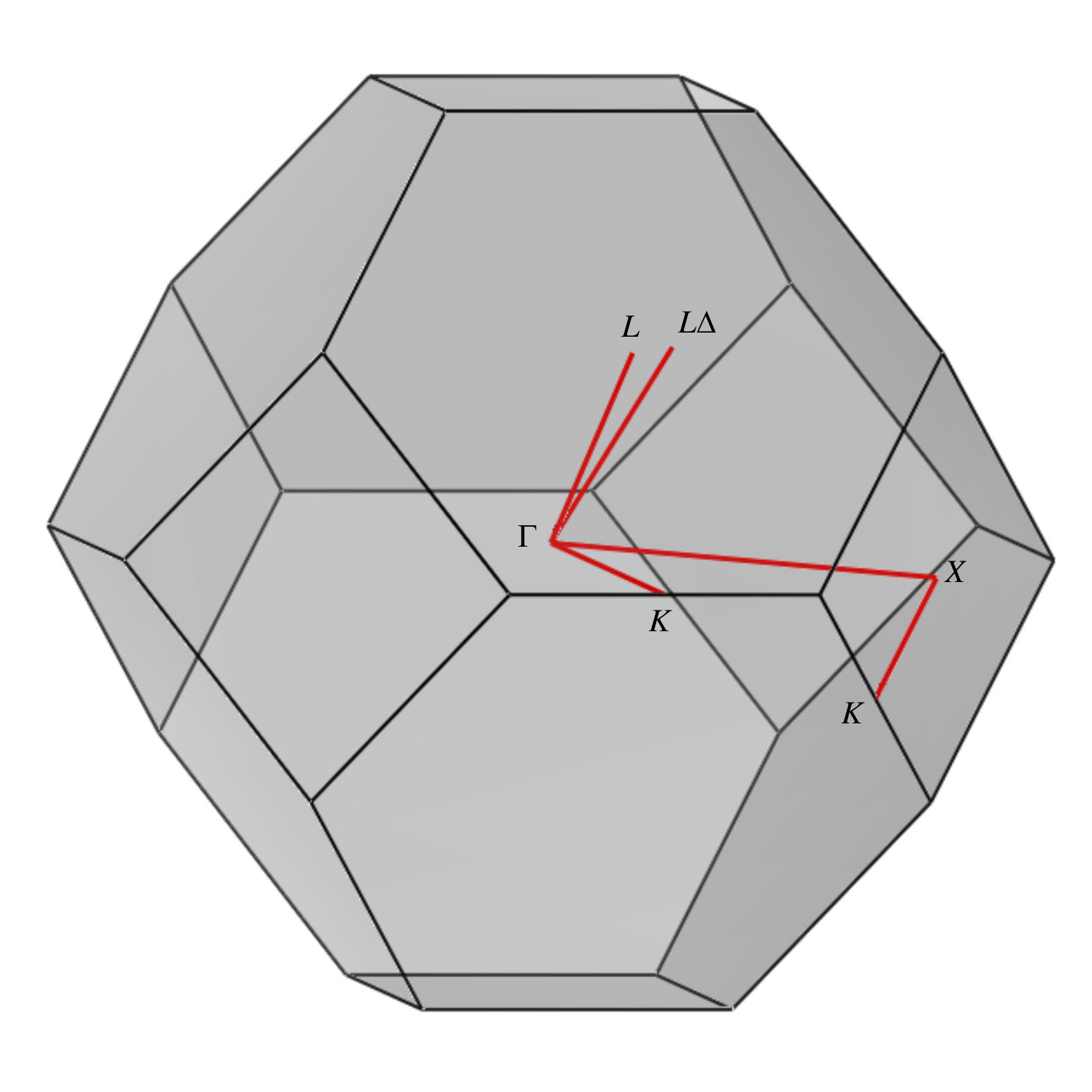}
\caption{\label{GeBZ} First Brillouin zone of Ge. The red lines show directions of interest. Important to notice is that the $L\Delta$ phonons, generated by the $L\rightarrow\Delta$ transition, occur very close to the $L$ point in the Brillouin zone. }
\end{figure}
\begin{figure}
\includegraphics[width=3.2in] {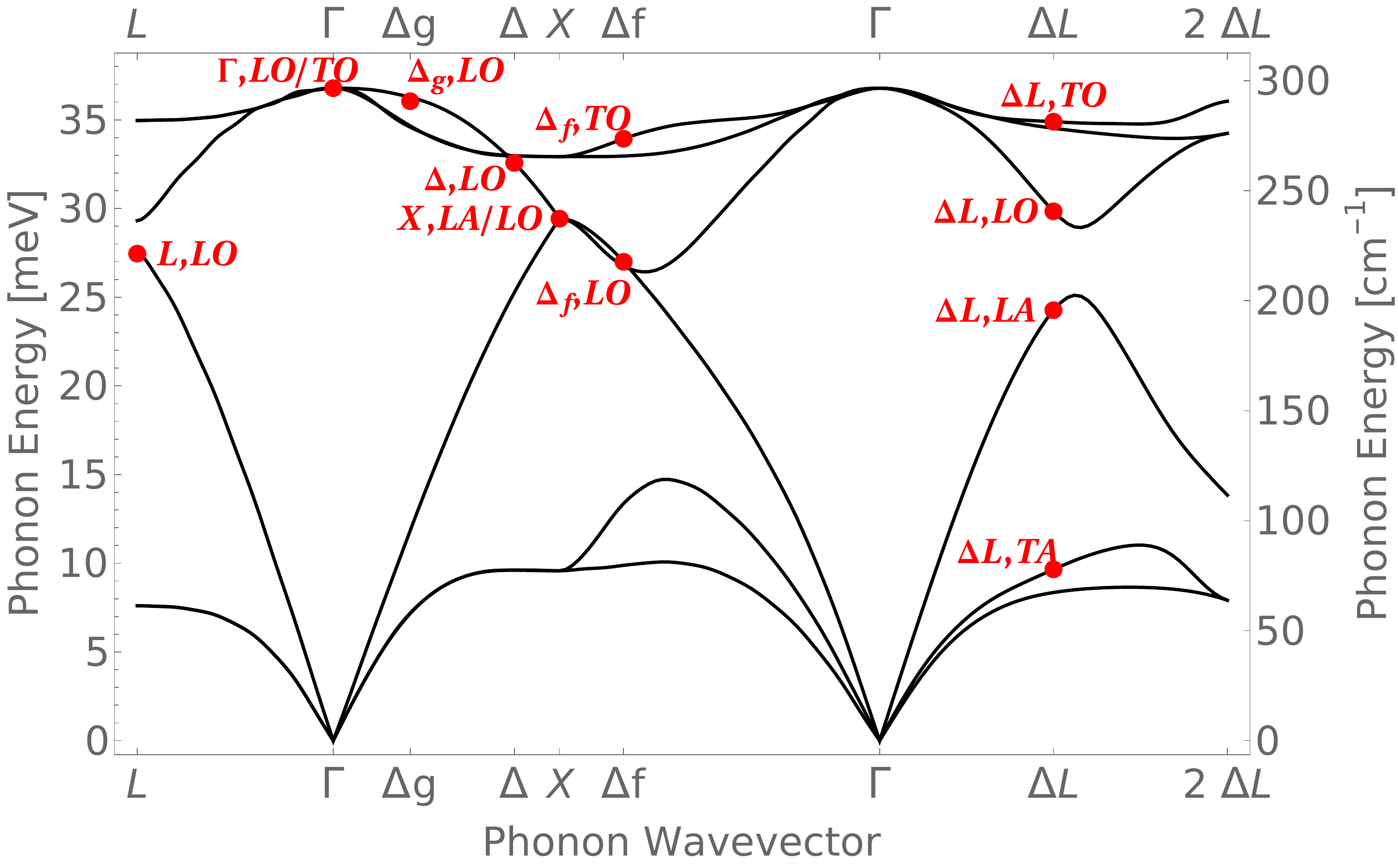}
\caption{\label{phondis} Phonon dispersion of Ge. The phonons generated by EPC are shown in red circles, and correspond to those shown in Table \ref{param}.}
\end{figure}

The phonons involved in inter-valley scattering, their wave-vectors, EPC matrix elements that generate them and anharmonic relaxation time are shown in Table \ref{param}. Fig. \ref{phondis} shows the phonon dispersion of Ge, with the phonons generated by inter-valley EPC marked by red circles. The labeling of the phonons is as follows: The $L$, $X$ and $\Gamma$ phonons correspond to phonons in the points of the Brillouin zone with the same name, and are generated by the transitions $\Gamma\rightarrow L$, $L\rightarrow L'$ and $L\rightarrow L$, respectively. The $\Delta$ phonon corresponds to a phonon at 0.83 in the $\Gamma-X$ line, and is generated by the $\Gamma\rightarrow \Delta$. As the remaining phonons do not correspond to any symmetry lines, they are labeled by the transitions that generate them. Thus, $L\Delta$ (see Fig. \ref{GeBZ}), $\Delta_g$, $\Delta_f$ are phonons that lie at the q-points equal to the difference of the $L$ and $\Delta$ points, two $\Delta$ points along the same Cartesian axis and two $\Delta$ at perpendicular axes, respectively.

\begin{table}
\caption{\label{param} List of the phonons generated by EPC, their calculated EPC scattering matrix element\cite{prb2,jap,jap2,prberratum} and phonon-phonon anharmonic decay times. Phonon wave-vectors are in reduced coordinates.}

\begin{tabular}{ l c c c c r }
\hline
\hline
Ph & Valleys $i,j$&Pol $\eta$ &Wave-vector $\bf q_0$&$D^{\eta,i,j}_{\bf q_0}$ &$\tau_{\bf q_0}^\eta$\\
&&&&(eV/\AA)&(ps)
\\
\hline
$L$&$\Gamma,L$&$LA$&$\left(\frac{1}{2},\frac{1}{2},\frac{1}{2}\right)$&4.12&17.4\\
$X$&$L,L$&$LA$&$\left(0,\frac{1}{2},\frac{1}{2}\right)$& 0.19&44.2\\
$X$&$L,L$&$LO$&$\left(0,\frac{1}{2},\frac{1}{2}\right)$& 1.11&44.2\\
$\Gamma$&$L,L$&$LO$/$TO$&$\left(0,0,0\right)$&2.97&2.2\\
$\Delta$&$\Gamma,\Delta$&$LO$&$\left(0,0.415,0.415\right)$& 2.50&12.3\\
$L\Delta$&$L,\Delta$&$TA$&$\left(0.5,0.085,0.085\right)$& 1.28& 29.8\\
$L\Delta$&$L,\Delta$&$LA$&$\left(0.5,0.085,0.085\right)$&2.55& 7.3\\
$L\Delta$&$L,\Delta$&$LO$&$\left(0.5,0.085,0.085\right)$&2.85&12.9\\
$L\Delta$&$L,\Delta$&$TO$&$\left(0.5,0.085,0.085\right)$&1.35& 3.0\\
$\Delta_g$&$\Delta_x,\Delta_x$&$LO$&$\left(0,0.17,0.17\right)$& 5.43&3.68\\
$\Delta_f$&$\Delta_x,\Delta_y$&$LO$&$\left(0.415,-0.415,0\right)$&2.57& 7.29\\
$\Delta_f$&$\Delta_x,\Delta_y$&$TO$&$\left(0.415,-0.415,0\right)$& 5.22& 4.2\\
\hline
\end{tabular}

\end{table}

Once the scattering rates have been determined, the rate of change in plasma temperature at each time step can be obtained by
%
\begin{eqnarray}
\frac{dT_p}{dt}(T,\mu)&=&\sum_{\bf i,f} \frac{1}{c_v(T,\mu)}\left[\left(R_{if}^--R_{if}^++R_{fi}^--R_{fi}^+\right)\hbar \omega_{if}+\right.\nonumber \\
&+&\left. \left(\frac{U_{i}}{n_i}-\frac{U_{f}}{n_f}\right)\left(R_{if}^-+R_{if}^+-R_{fi}^--R_{fi}^+\right)\right]
\end{eqnarray}
where the electron heat capacity is given by $c_v=\sum_i\partial U_i/\partial T_p$, the energy $U_i=\int dE g_i(E) E f_i(E)$ and $g_i(E)$ the electronic density of states of valley $i$.

We set the initial phonon population and $\Delta$ and $L$ valley carrier density at their equilibrium values for intrinsic Ge at 300 K. Conversely, the non-equilibrium $\Gamma$ conduction band valley density corresponds to that generated by population inversion with a 1.5 eV 50 fs light pulse across the direct band gap at $\Gamma$ of 0.9 eV. We estimate an initial photoexcited carrier population of $n_{\Gamma}=10^{20} cm^{-3}$. We estimate an initial plasma temperature and for the $\Gamma$ valley of $T_p=1160.5K$, with the resulting chemical potential $\mu_\Gamma=0.3525$eV$+E_L$, respectively, where $E_L=0.67$eV is the energy of the bottom of the conduction band valley at $L$ relative to the top of the valence band. The chemical potentials for the $\Delta$ and $L$ valleys at this plasma temperature and intrinsic carrier concentration results in $\mu_{\Delta,L}=E_L-0.3$eV. 

\subsection{Phonon-phonon scattering}\label{sphph}
As phonons of a particular mode and wave-vector become populated via
electron-phonon scattering as described above, phonon-phonon scattering will
cause this energy to be distributed to other phonons in the system.

This is the same process that is considered when calculating the lattice
contribution to the thermal conductivity. In lattice thermal conductivity
calculations it is typically found that truncating the expansion of the energy
in terms of the atomic positions, i.e. including only three-phonon scattering
processes yields to calculated values of the thermal conductivity in good
agreement with experiment. In our calculations, we have modified an in-house
code we previously employed successfully to obtain the lattice thermal
conductivity of several materials including PbTe and Pb$_{1-x}$Ge$_x$Te
alloys,\cite{murphy2016, murphy2017} to calculate the time
evolution of phonon populations throughout the Brillouin zone coupled through
these phonon-phonon interactions.

Restricting the simulation to include only three-phonon processes means that
for a given phonon there are two types of phonon-phonon scattering event that
will lead to a change in its population: (1) the phonon scattering with a second
phonon to create a third, and (2) the phonon scattering to create two other
phonons. We term these class 1 and class 2 events respectively. Along with
this we must also include the creation and annihilation of phonons as
electrons scatter between different valleys as outlined above.

We wish to model a system of interacting phonons, with a momentum ${\bf q}_i$ and
mode $\eta_j$. Following Ref.~\onlinecite[Section~6.4.1]{srivastava}, we consider an initial
state of three phonons in the system as $\left| i \right> \equiv 
\left|N_{q_1 \eta_1}, N_{q_2 \eta_2}, N_{q_3 \eta_3}\right>$. An anharmonic
perturbation
$\mathcal{V}_3$ causes the system to change in time $t$ to a final state
$\left| f \right>$:
\textit{for class 1 events} $\left|f\right> = \left| N_{q_1 \eta_1} - 1, N_{q_2
\eta_2} - 1, N_{q_3 \eta_3} + 1 \right>$, and  \textit{for class 2 events}
$\left|f\right> = \left| N_{q_1 \eta_1} - 1, N_{q_2 \eta_2} + 1, N_{q_3 \eta_3} + 1
\right>$.
The transition probability is given by Fermi's golden rule:
\begin{equation}\label{eq:FGR}
  P_{i}^{f} = \frac{2 \pi}{\hbar} \left|\braket{f}{\mathcal{V}_3}{i}\right|^2
  \delta(E_f - E_i)
\end{equation}
where $E_i$ and $E_f$ are the energies of the initial and final three-phonon
systems.
For a class 1 event, Eq.~\ref{eq:FGR} becomes
\begin{align}
  P_{q_1 \eta_1, q_2 \eta_2}^{q_3 \eta_3} =& \frac{2 \pi}{\hbar^2}
  \left|
  \left< N_{q_1 \eta_1} - 1, N_{q_2 \eta_2} - 1, N_{q_3 \eta_3} + 1\right|\right. \nonumber\\
  & \left.\mathcal{V}_3 \left|
  N_{q_1 \eta_1} N_{q_2 \eta_2} N_{q_3 \eta_3}\right> \right|^2 \nonumber\\
  &\times \delta (\omega_{q_3 \eta_3} - \omega_{q_2 \eta_2} - \omega_{q_1 \eta_1})
\end{align}

We can reformulate this as
\begin{align}\label{eq:RC1}
  P_{q_1 \eta_1, q_2 \eta_2}^{q_3 \eta_3}  = &
  \mathcal{R}_{123} N_{q_1 \eta_1} N_{q_2 \eta_2} (N_{q_3 \eta_3} + 1)
  \delta_{q_1+q_2+q_3,G}\nonumber\\
   & \times \delta(e_{q_3 \eta_3} - e_{q_2 \eta_2} - 
  e_{q_1 \eta_1}),
\end{align}
where $e=\hbar\omega$, the kernel $\mathcal{R}_{123}$ is related to the third
order interaction between the modes labeled 1, 2 and 3, and contains all the
other factors not expressed in Eq.~\ref{eq:RC1}. The various $n$ and $n+1$ 
factors arise through the normalization of the creation and annihilation
operators.
Similarly, class 2 events can be formulated as
\begin{align}\label{eq:RC2}
  P_{q_1 \eta_1}^{q_2 \eta_2, q_3 \eta_3}  = &
  \mathcal{R}_{123} N_{q_1 \eta_1} (N_{q_2 \eta_2} + 1) (N_{q_3 \eta_3} + 1)
  \delta_{q_1+q_2+q_3,G}\nonumber\\
   & \times \delta(e_{q_1 \eta_1} - e_{q_2 \eta_2} - e_{q_3 \eta_3})
\end{align}
As only the phonon populations $N_{q, s}$ evolve during the
simulation, the various $\mathcal{R}_{123} \delta_{q_1+q_2+q_3,G} \delta(e_{q_3 \eta_3} - e_{q_2 \eta_2} - 
  e_{q_1 \eta_1})$ can be precomputed.

The total rate of change of the population of the phonon with wave-vector $q_1$
and mode $\eta_1$ due to phonon-phonon scattering can be found by summing the
various $P$ terms which involve this phonon as
\begin{align}
  \label{eq:dndt3}
  \frac{\partial N_{q_1 \eta_1}}{\partial t}  = \frac{1}{2}
  \displaystyle\sum_{q_2 \eta_2, q_3 \eta_3} & [P_{q_3 \eta_3}^{q_1 \eta_1, q_2 \eta_2} - 
    P_{q_1 \eta_1, q_2, \eta_2}^{q_3 \eta_3}\nonumber \\
    & + P_{q_2 \eta_2}^{q_1 \eta_1, q_3 \eta_3} - P_{q_1 \eta_1, q_3 \eta_3}^{q_2 \eta_2}
    \nonumber \\
    & +P_{q_2 \eta_2, q_3 \eta_3}^{q_1 \eta_1} - P_{q_1 \eta_1}^{q_2 \eta_2, q_3 \eta_3}].
\end{align}
with the additional factor of $\frac{1}{2}$ to correct for the double counting
of each of the $P$ terms due to the interchange of the phonons referred to by
$q_2 \eta_2$ and $q_3 \eta_3$.

Otherwise the approach used follows the linear tetrahedron method as described
in Ref.~\onlinecite{blochl} (we do not use the additional nonlinear
corrections described in this paper), but with the weightings modified
as described in Appendix~\ref{ap:econs} to conserve
energy by construction. We use this approach to transform the integral over
q-points to a weighted sum. For each mode at each q-point
a grid is constructed for each possible pair of interacting
phonons such that momentum is conserved. That is, for the mode each phonon
labeled by wave-vector and mode index $(q_1, \eta_1)$ 
3 grids are constructed where the values at each point and corresponding
scatterings are
\begin{enumerate}
\item $e(q_3, \eta_3) - e(q_2, \eta_2) - e(q_1, \eta_1),$
\item $e(q_2, \eta_2) - e(q_3, \eta_3) - e(q_1, \eta_1),$
\item $e(q_2, \eta_2) + e(q_3, \eta_3) - e(q_1, \eta_1),$
\end{enumerate}
and where $q_2$ is any wave-vector on the simulated grid, and $q_3$ is set by
conservation of momentum. A weighting is calculated for
each point on the grid using tetrahedra. These are combined with the 
scattering rate kernel and the population factors to determine the rate
at which each of the three modes involved is changing population. 
After these are fully summed up, the $dn/dt$ terms are divided by an
additional factor of three relative to Eq.~\ref{eq:dndt3} above
as we have tracked the changes in all 3 involved modes and
explicitly triple counted.

We use third order force constants calculated using finite displacements in
64-atom supercells, generated using the \textsc{Phono3py} code\cite{phono3py},
and forces calculated using the \textsc{abinit} code\cite{abinit} to simulate
the evolution of the phonon populations on a $20\times20\times20$ grid of
$q$-points. The phonon populations absorbed and emitted due to electron-phonon
scattering are pre-calculated on the same grid and incorporated directly into
the simulation.

\section{Conclusions}
We have calculated the time evolution of the phonon and electron distributions and x-ray diffuse scattering intensity using coupled Boltzmann rate equations and parameters calculated entirely from first-principles. We have compared our calculations to measurements of the time-resolved diffuse x-ray scattering intensity in bulk Ge during the first 10ps after photo-excitation with a 50fs infra-red laser pulse centered at 800nm, at the LCLS free-electron laser facility. Our calculations and experiment are in excellent agreement. In particular, we explain the slow increase in x-ray intensity near the $L$ point in the BZ within the first 10ps after photo-excitation, as due entirely to inter-valley electron-phonon scattering between the $L$ and $\Delta$ valleys. In particular, we are able to explain the coffee bean shape of the increase in intensity at the $L$ point as due to the distribution of carriers at the $L$ and $\Delta$ valleys and the polarization of the EPC generated phonons. 
We also account for the faint diffuse glow throughout the Brillouin zone as due to anharmonic phonon-phonon scattering of the phonon distribution generated by electron-phonon scattering.

The theoretical results yield a richer non-equilibrium phonon distribution than is possible to measure in the current experimental configuration. This suggests that changing the observed portion of the Brillouin zone by modifying the x-ray incidence angles and energies would show the dynamics of all other phonons involved in the relaxation process.

In this work we purposely ignored the effects of acoustic long wavelength phonons generated by intra-valley electron-phonon scattering. Including these, at the cost of denser phonon grids, would yield further understanding of the coherent phonon effects near the zone center observed in the first ps after photoexcitation, and the transition to incoherent phonon distributions.

Finally, these diffuse x-ray scattering intensity increase produced by inter-valley electron-phonon scattering should be observable in many other many-valley semiconductors, such as GaAs and PbTe.



\section{Acknowledgments}
FM-A was supported by Science Foundation Ireland project 12/IA/1601. MT and DAR were supported by the U.S. Department of Energy, Office of Science, Office of Basic Energy Sciences through the Division of Materials Sciences and Engineering under Contract No. DE-AC02-76SF00515. Use of the Linac Coherent Light Source (LCLS), and the Stanford Synchrotron Radiation Lightsource at SLAC National Accelerator Laboratory, is supported by the U.S. Department of Energy, Office of Science, Office of Basic Energy Sciences under Contract No. DE-AC02-76SF00515.

\appendix
\section{Evaluation of scattering rate integrals retaining energy conservation}
\label{ap:int}
In this appendix we explain the details of how the integral in Eq. \ref{rq} is evaluated.
We need to evaluate
\begin{equation}
\int_{BZ} {\bf dk} f_{\bf k}\left(1- f_{\bf k+q}\right) \delta(h({\bf k,q}))
\end{equation}
where $h({\bf k,q})=E_{\bf k+q}-E_{\bf k}\pm\hbar \omega^\eta_{\bf q_0}$. The standard way of evaluating this delta is by a change of variables from ${\bf dk}\rightarrow {\bf dS} dh$, where $\bf dS$ is the surface in $k$-space at constant $h$, i.e. perpendicular to $dh$. This gives
\begin{equation}
\int_{BZ} {\bf dS}  dh f_{\bf k} \left(1- f_{\bf k+q}\right) \frac{\delta(h)}{\left|\nabla_{\bf k} h\right|}.
\end{equation}
Integrating in $h$ 
\begin{equation}\label{intfh}
\left. \int_{BZ} {\bf dS}  \frac{f_{\bf k} \left(1- f_{\bf k+q}\right)}{\left|\nabla_{\bf k} h\right|}\right|_{{\bf k} | h=0},
\end{equation}
where the integration is over the surface in $k$-space that satisfies $h=0$. 
Finding the surfaces in $k$-space that satisfy $h=0$ is greatly simplified by having analytic expressions for the energy dispersion of the valleys. We therefore parametrize the $\Delta$ and $L$ valley dispersions using parabolic functions, and the highly non-parabolic $\Gamma$ valley with quartic functions in $k_x$, $k_y$ and $k_z$. 

\begin{figure}
\includegraphics[width=3.2in] {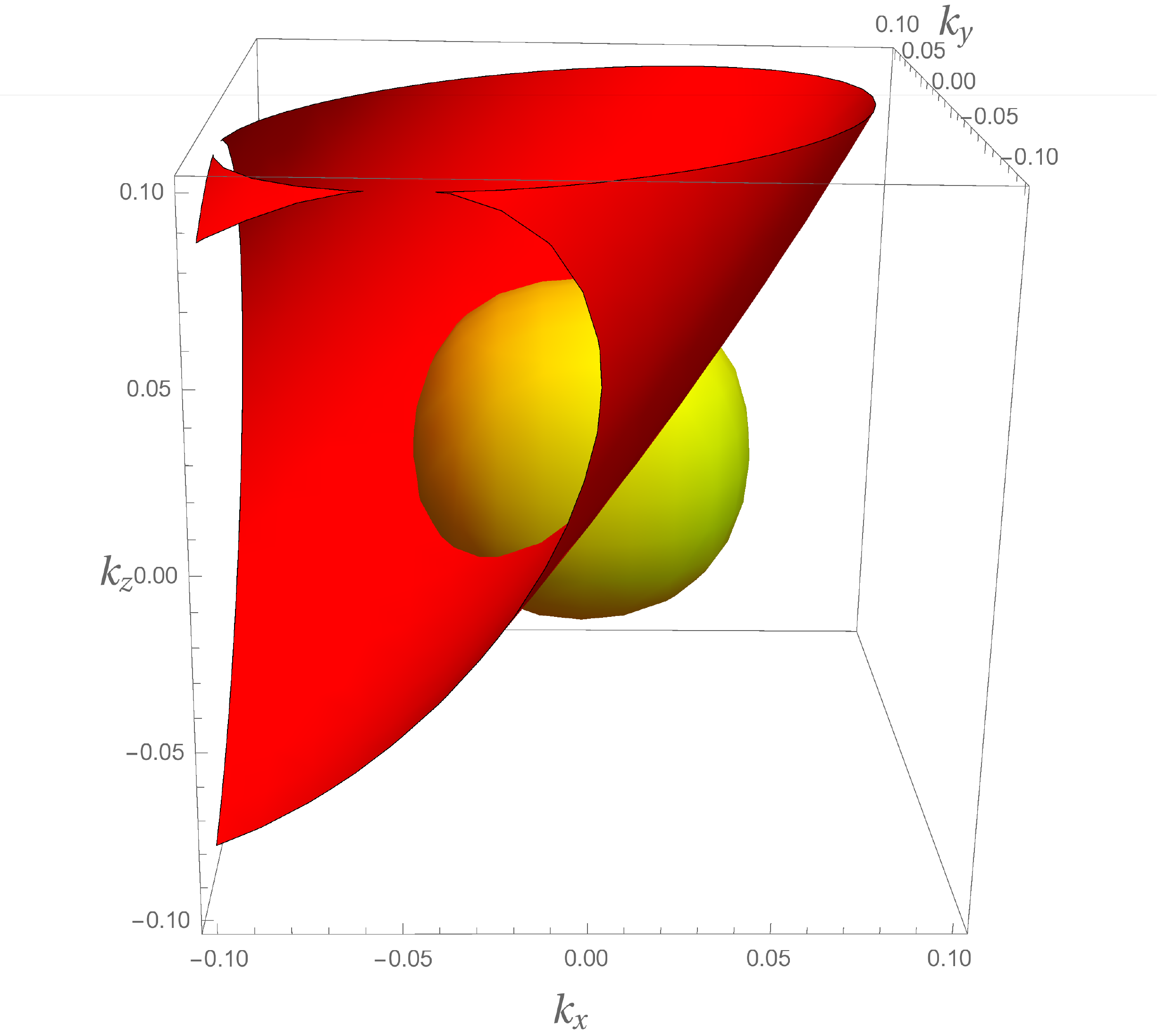}
\caption{\label{glqpq} Iso-surfaces at energy 0.3 eV above $E_L$ of the $\Gamma$ valley (yellow, center) and the $L$ valley (red) translated by wave-vector ${\bf q}=\frac{2\pi}{a}(0.505,0.505,0.495)$.}
\end{figure}

\begin{figure}
\includegraphics[width=3.2in] {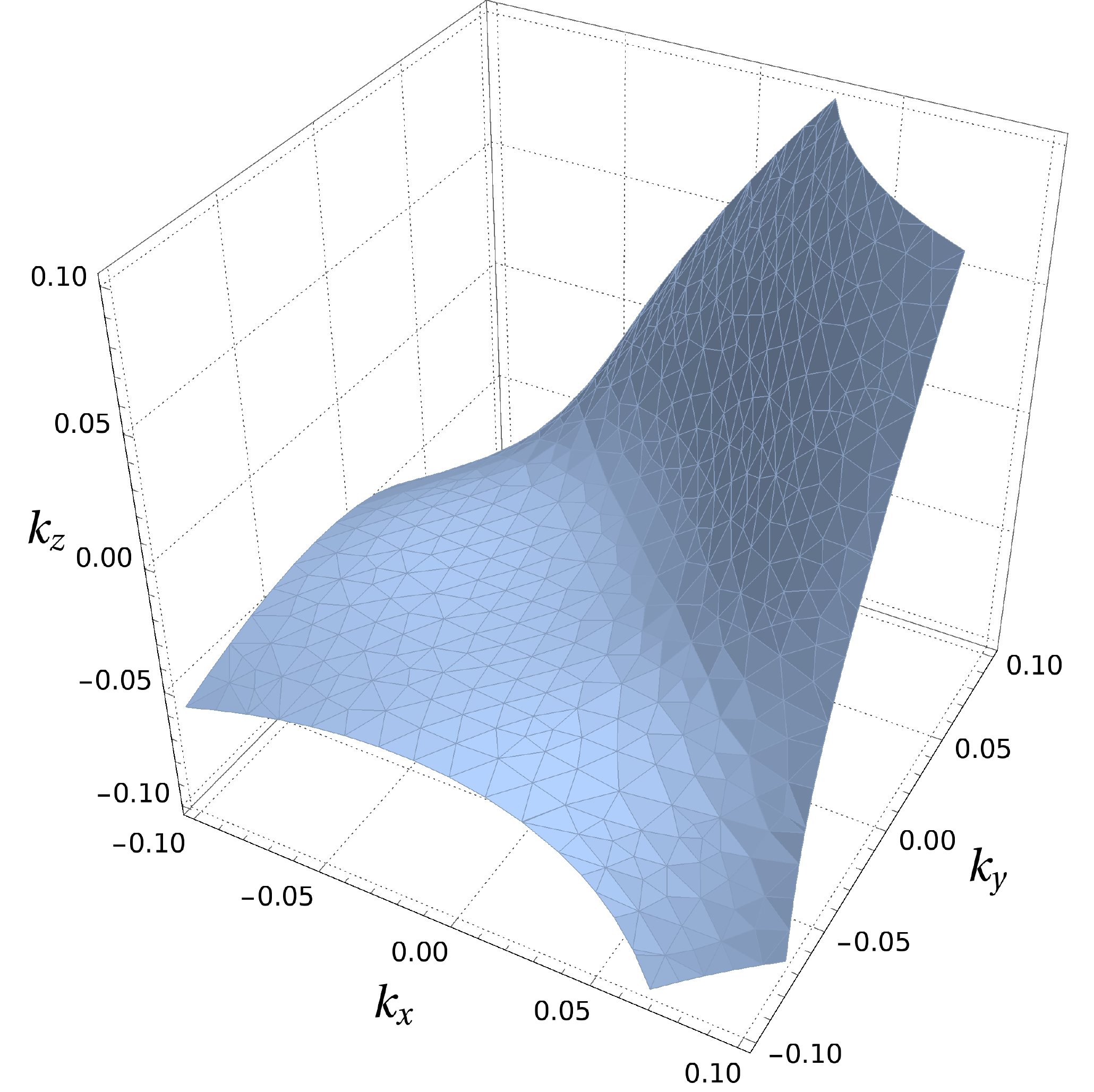}
\caption{\label{sglq} Surface $S({\bf k})$ with $h({\bf k},{\bf q})=0$ at fixed ${\bf q}=\frac{2\pi}{a}(0.505,0.505,0.495)$.}
\end{figure}

When involving inter-valley scattering, the evaluation of integral \ref{intfh} requires a very dense set of points in a very small portion of the Brillouin Zone. This can be better illustrated by an example. Consider inter-valley scattering between the $\Gamma$  and $L$ valleys. The surface in $k$-space at which $h=0$ for a given $\bf q$ is obtained by the intersection of the energy iso-surfaces of the two valleys, when one of them is translated by $\bf q$, as shown in Fig. \ref{glqpq} for $q=\frac{2\pi}{a}\left(0.505,0.505,0.495\right)$. This generates the surface $S({\bf k})|_{h({\bf k,q})=0}$ shown in Fig. \ref{sglq}, when performed for all energy iso-surfaces.
\begin{figure}
\includegraphics[width=3.2in] {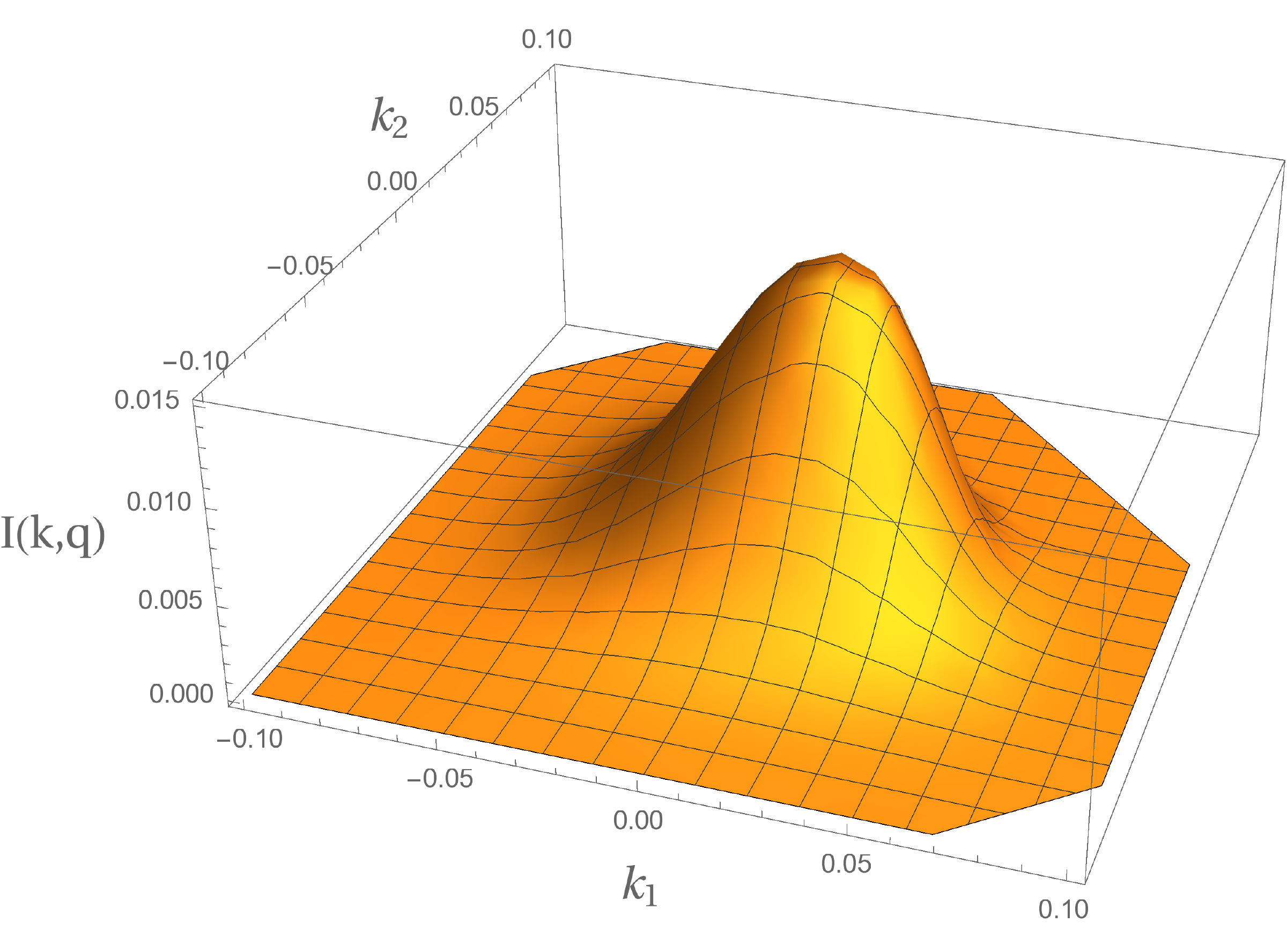}
\caption{\label{ingrnd} Integrand of Eq. \ref{intfh} on surface $S({\bf k})$ of $h({\bf k},{\bf q})=0$ with ${\bf q}=\frac{2\pi}{a}(0.505,0.505,0.495)$.}
\end{figure}

\begin{figure}
\includegraphics[width=3.2in] {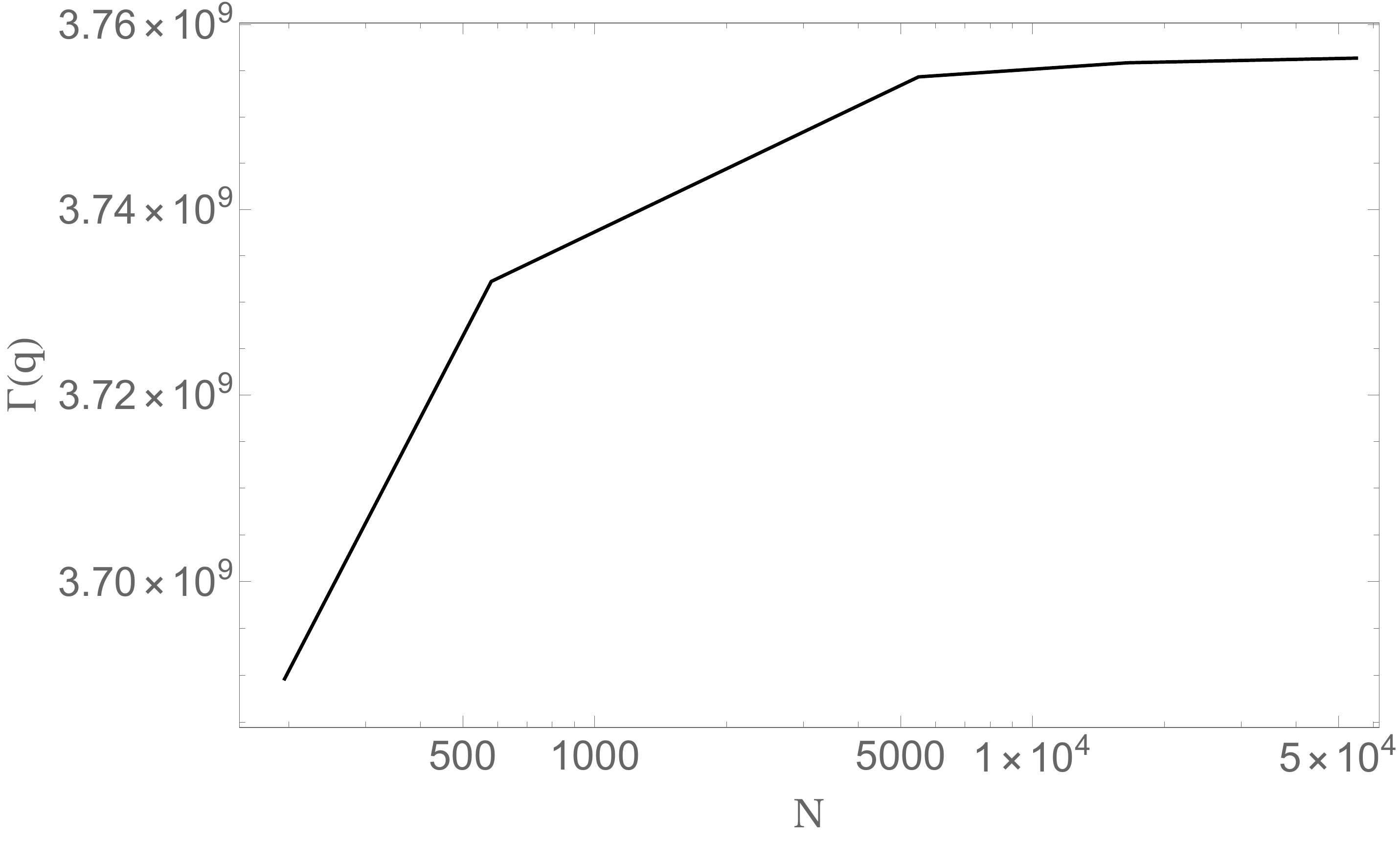}
\caption{\label{gammaqvsn} Convergence of integral in Eq. \ref{intfh} vs number of points in the integrand surface mesh.}
\end{figure}

\begin{figure}
\includegraphics[width=3.2in] {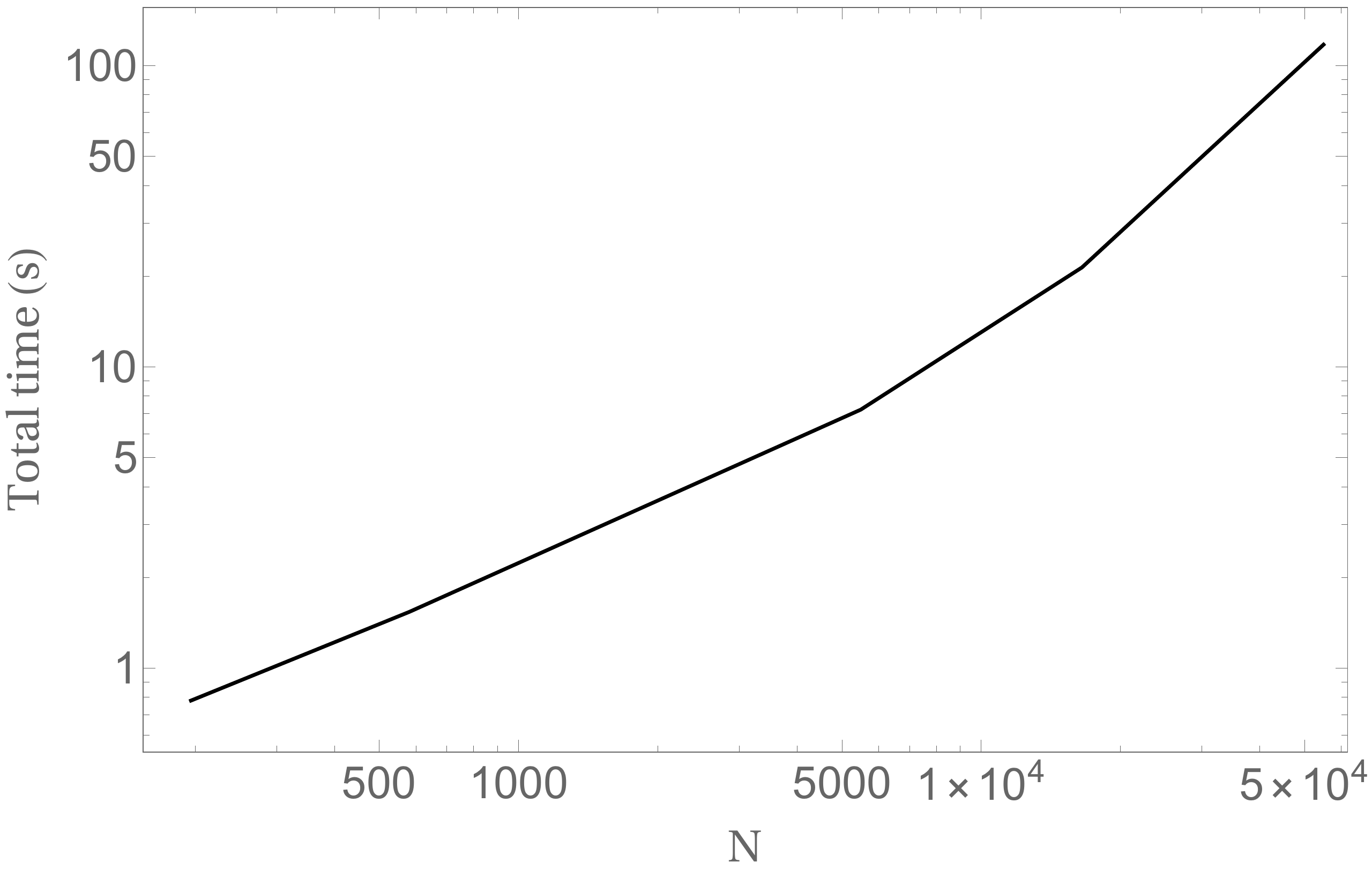}
\caption{\label{gammaqvsnvst} Time required to integrate Eq. \ref{intfh} vs number of points in the integrand surface mesh.}
\end{figure}

\begin{figure}
\includegraphics[width=3.2in] {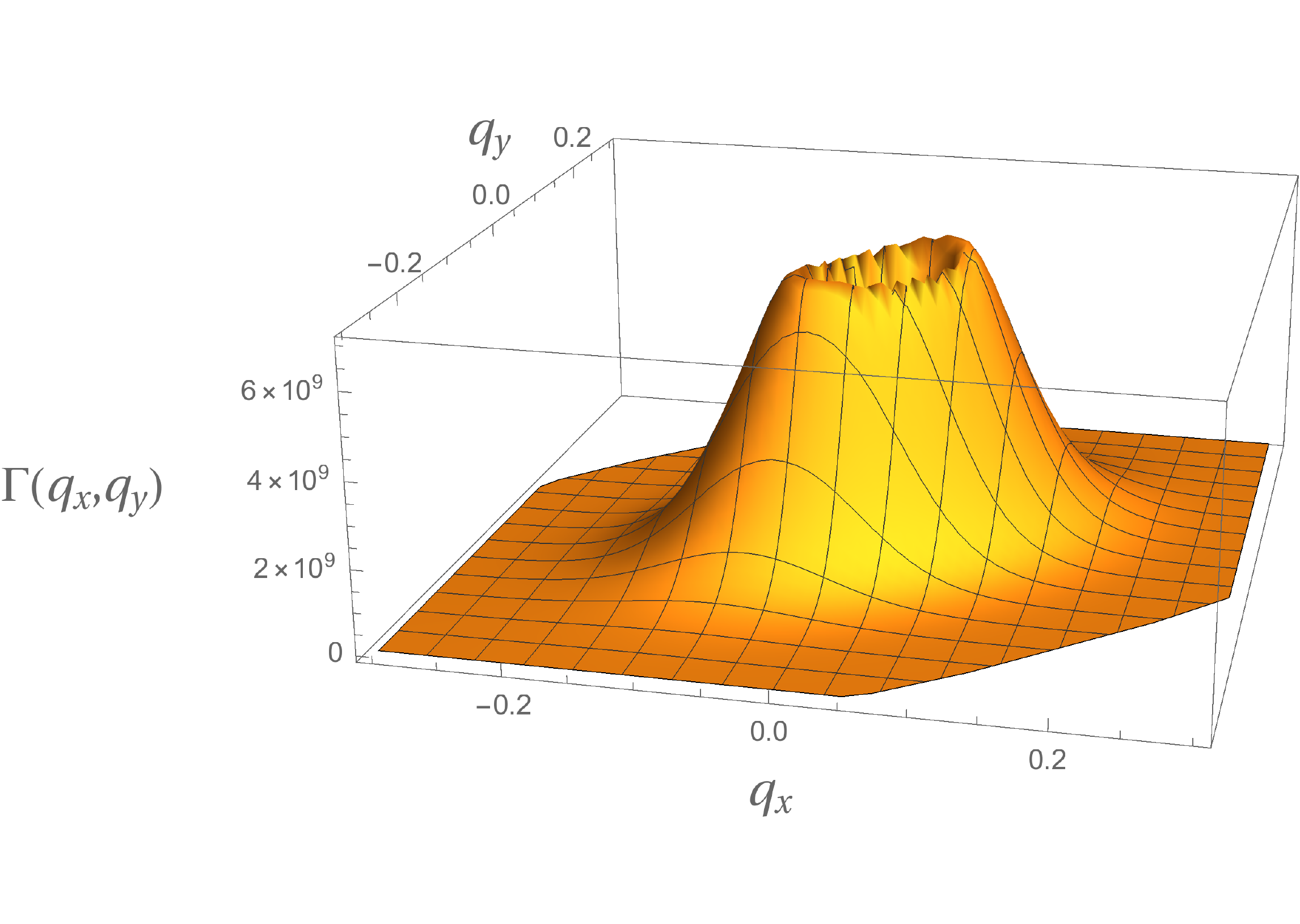}
\caption{\label{gammaqxqy} Phonon generation rate $\Gamma({\bf q}-{\bf q}_0)$ near ${\bf q}_0={\bf q}_L$, for $q_z-{\bf q}_{0,z}=0.01 \frac{2\pi}{a}$.}
\end{figure}

 The integrand of Eq. \ref{intfh} is shown in Fig. \ref{ingrnd} for ${\bf q}=\frac{2\pi}{a}(0.505,0.505,0.495)$. The evaluation of this integrand is performed on the tessellated mesh of $S({\bf k,q})$ shown in Fig. \ref{sglq}. The convergence of the integral of $\Gamma({\bf q})$ with the number of points on the mesh is shown in Fig. \ref{gammaqvsn}. A sample of the time vs sampling point density is shown in Fig. \ref{gammaqvsnvst}, showing an exponential growth of clock-time vs number of points. The resulting EPC generation rate $\Gamma({\bf q})$ for this example is shown in Fig. \ref{gammaqxqy} at fixed $q_z$.

The phonon generation rate $\Gamma({\bf q})$ serves two separate functions, that require different $q$-point grids. 
The first is to update the phonon populations at each time step in the $20^3$ grid that generates the diffuse x-ray scattering intensity. 
In practice, $\Gamma({\bf q})$ decays rapidly away from the ${\bf q}_0$ of Table \ref{param} (see Fig. \ref{gammaqxqy}). 
Hence we only calculate $\Gamma({\bf q})$ on points of the 20$^3$ grid near  the ${\bf q}_0$. 
The second function is the integral of the inter-valley electron-phonon scattering rate, Eq. \ref{rij}. 
This integral needs a denser grid than the 20$^3$ around the ${\bf q}_0$ in a small region of the Brillouin zone (Fig. \ref{gammaqxqy}) and is performed numerically using global adaptive grids \cite{math,math2}.

\section{Energy conservation in Phonon-Phonon Scattering Simulation}
\label{ap:econs}
An important feature of a simulation of the time evolution of phonon
populations due to phonon-phonon scattering is that it conserves
energy. However, this is not guaranteed in a straightforward integration
of the rates of change in populations, due to
the need to represent the phonon populations on a finite grid, and
approximate the $\delta$ function for energy conservation somehow, with
Gaussian broadening or linear tetrahedron integration commonly used in e.g.
lattice thermal conductivity calculations. To see how this can lead to
an issue, consider a case where we have three interacting modes:
$\alpha$, $\beta$, and $\gamma$, with
 populations $n_\alpha$, $n_\beta$, and $n_\gamma$, and energies $e_\alpha$,
 $e_\beta$, and $e_\gamma$, with $e_\alpha \approx e_\beta + e_\gamma$,
 such that $e_\alpha - e_\beta - e_\gamma = \epsilon$ where $\epsilon$ is
 small compared to any of the individual energies.
 Say we approximate the $\delta$ function term as $\Delta$:
\begin{equation}
  \delta(e_\alpha - e_\beta - e_\gamma) \approx \Delta.
\end{equation}
Our expressions for the rates of change of populations become
\begin{align}
  \label{eq:dn123a}
  \frac{\partial N_\alpha}{\partial t} = &P_{\beta \gamma}^{\alpha} -
  P_{\alpha}^{\beta \gamma} \nonumber\\
  = & \mathcal{R}_{123} [(N_\alpha + 1) N_\beta N_\gamma - 
  N_\alpha(N_\beta+1)(N_\gamma+1)]\Delta,
\end{align}
\begin{align}
  \label{eq:dn123}
  \frac{\partial N_{\beta}}{\partial t} = &P_{\alpha}^{\beta \gamma} -
  P_{\beta \gamma}^{\alpha} \nonumber\\
  = & \mathcal{R}_{123} [N_\alpha(N_\beta+1)(N_\gamma+1) - 
  (N_\alpha+1)N_\beta N_\gamma]\Delta \nonumber\\
  = & \frac{\partial N_{\gamma}}{\partial t} = 
  - \frac{\partial N_{\alpha}}{\partial t}
\end{align}

The initial vibrational energy $V$ of the system can be written as
\begin{equation}
  N_\alpha e_\alpha+ N_\beta e_\beta+ N_\gamma e_\gamma=V.
\end{equation}
For energy to be conserved in the scattering we need
\begin{align}
  \label{eq:econs}
  \frac{\partial V}{\partial t}
  = & e_\alpha \frac{\partial N_\alpha}{\partial t} + 
  e_\beta \frac{\partial N_\beta}{\partial t} + e_\gamma
  \frac{\partial N_\gamma}{\partial t}
  \nonumber \\
  = & 0
\end{align}
Combining this with Eq.~\ref{eq:dn123} above we have
\begin{align}
  \frac{\partial V}{\partial t}
  = & e_\alpha \frac{\partial N_\alpha}{\partial t} - 
  e_\beta \frac{\partial N_\alpha}{\partial t} - e_\gamma
  \frac{\partial N_\alpha}{\partial t}
  \nonumber \\
  = &\frac{\partial N_\alpha}{\partial t} \left(e_\alpha - e_\beta -
  e_\gamma\right),
\end{align}
showing that the total energy will only be conserved if
$e_\alpha - e_\beta -e_\gamma = 0$, which if we we simulate our system on a
grid of wave-vectors, can only be achieved in the limit
of an infinitely dense sampling.

To work around this we use a modified linear tetrahedron approach to evaluate
the integration over the $\delta$-function in Eq.~\ref{eq:RC1} and
\ref{eq:RC2}. We decompose the grid of $q$-points into tetrahedron,
approximating the two phonon density of states linearly within each one.
Now instead of simulating the scattering between a fixed grid of points
in the Brillouin zone, we can consider scattering to and from surfaces
satisfying the delta-function condition above that intersect through
tetrahedra. Then the assignment of the area associated with intersecting
the surface to the grid points composing the tetrahedron can be constructed
so as to ensure the energy is conserved in this process.

We can see this by expanding the example above, where we have three modes
which scatter among each other and now looking at how this behaves over a
tetrahedron of 4 groups of q-points, with momentum conserved for the
scattering process at each group of points. Now we have a set of energies
$e_{u,i}$ and populations $N_{u,i}$ with $u$ indexing one of the three modes
$\alpha$, $\beta$ and $\gamma$, and $i=1$ to $4$ indexing the corners of the
tetrahedron. A single scattering kernel $\mathcal{R}_{123}$ is then
assigned to the tetrahedron. In our implementation we do this by taking a
weighted average over the matrix element from each corner of the tetrahedron
(how these weightings are calculated is discussed later in this section).
Similarly, a single population for each mode is assigned to the tetrahedron;
with the same weighted averaging scheme used in our implementation.

Equations \ref{eq:dn123a} and \ref{eq:dn123} above still hold, but now refer
to the scattering kernel and populations on a tetrahedron. 
The total
vibrational energy of the system becomes
\begin{equation}
  \Sigma_i w_i (N_\alpha e_{\alpha, i}+ 
  N_\beta  e_{\beta, i} +
  N_\gamma e_{\gamma,i}) =V.
\end{equation}
For energy to be conserved in the scattering we now need
\begin{align}
  \frac{\partial V}{\partial t}
  = & \Sigma_i w_i \left(e_{\alpha,i} \frac{\partial N_\alpha}{\partial t} + 
  e_{\beta,i} \frac{\partial N_\beta}{\partial t} + e_{\gamma,i}
  \frac{\partial N_\gamma}{\partial t}\right)
  \nonumber \\
  = & 0.
\end{align}

Again we can use Eq.~\ref{eq:dn123} to rewrite the condition for energy
conservation as
\begin{equation}\label{eq:econs_tet}
\frac{\partial N_\alpha}{\partial t} \Sigma_i w_i 
(e_{\alpha, i} - e_{\beta, i} - e_{\gamma, i}) = 0.
\end{equation}
This can be enforced by an appropriate choice of $w_i$ within each
tetrahedron. The other feature that is desirable in choosing these weights
is that weights are larger for corners of the tetrahedron that are closer to
the surface intersecting it. The scheme we have used fulfills both of
these requirements. We first assign a delta energy value to each corner
of the tetrahedron as e.g. $e_i = e_{\alpha,i} - e_{\beta, i} - e_{\gamma, i}$
or whichever combination of energies is appropriate for the scattering process
we are considering. These are then sorted in increasing order of energy. The
surface we are calculating is that with $e_i=0$. If all $e_i$ are positive,
or all are negative, then this surface does not intersect the tetrahedron and
there will be no contribution. We then have three possible sets of weights
depending on how the surface intersects the tetrahedron.
For $e_1<0<e_2$ the normalized $w_i$ are
\begin{align}
w_1 = & \frac{1}{3}\left(3 + \frac{e_1}{e_2-e_1}+
  \frac{e_1}{e_3-e_1}+\frac{e_1}{e_4-e_1}\right)\nonumber\\
w_2 = & \frac{1}{3}\left(1 - \frac{e_2}{e_2-e_1}\right)\nonumber\\
w_3 = & \frac{1}{3}\left(1 - \frac{e_3}{e_3-e_1}\right)\nonumber\\
w_4 = & \frac{1}{3}\left(1 - \frac{e_4}{e_4-e_1}\right).
\end{align}
For $e_2<0<e_3$ the normalized $w_i$ are
\begin{align}
w_1 = & \frac{1}{4}\left(2 + \frac{e_1}{e_3-e_1}+
  \frac{e_1}{e_4-e_1}\right)\nonumber\\
w_2 = & \frac{1}{4}\left(2 + \frac{e_2}{e_3-e_2}+
  \frac{e_2}{e_4-e_2}\right)\nonumber\\
w_3 = & \frac{1}{4}\left(2 - \frac{e_3}{e_3-e_1}-
  \frac{e_3}{e_3-e_2}\right)\nonumber\\
w_4 = & \frac{1}{4}\left(2 - \frac{e_4}{e_4-e_1}-
  \frac{e_4}{e_4-e_2}\right).
\end{align}
For $e_3<0<e_4$ the normalized $w_i$ are
\begin{align}
w_1 = & \frac{1}{3}\left(1 + \frac{e_1}{e_4-e_1}\right)\nonumber\\
w_2 = & \frac{1}{3}\left(1 + \frac{e_2}{e_4-e_2}\right)\nonumber\\
w_3 = & \frac{1}{3}\left(1 + \frac{e_3}{e_4-e_3}\right)\nonumber\\
w_4 = & \frac{1}{3}\left(3 - \frac{e_4}{e_4-e_1}-
  \frac{e_4}{e_4-e_2}-\frac{e_4}{e_4-e_3}\right).
\end{align}
It can be shown that in each case these yield $\Sigma_i w_i e_i = 0$ such that
Eq.~\ref{eq:econs_tet} is satisfied and energy will be conserved in the
scattering simulation.

\bibliography{gexraymurphy_EM_181218}

\end{document}